\documentclass[a4paper,12pt,dvipsnames,usernames]{article}
\pdfoutput=1

\usepackage{aas_macros}

\usepackage{jcappub} 

\usepackage[utf8]{inputenc}
\usepackage{graphicx}
\usepackage{amsmath}
\usepackage{amssymb}
\usepackage{bm}
\usepackage{paralist,array}
\usepackage{units} 
\graphicspath{{./Plots/}}
\usepackage{slashed}
\usepackage{todonotes}
\usepackage[parfill]{parskip}
\usepackage{longtable}
\usepackage{subfig}

\graphicspath{{plots/}}

\newcommand{\eV}{\:\unit{eV}}
\newcommand{\cm}{\unit{cm}}

\newcommand{\CNB}{\ensuremath{{\mathrm{C}\nu\mathrm{B}}}}
\newcommand{\DR}{\ensuremath{{\mathrm{DR}}}}
\newcommand{\DM}{\ensuremath{{\mathrm{DM}}}}

\title{Probing sub-eV Dark Matter decays with PTOLEMY}
 
\author[1,2]{Kyrylo~Bondarenko,}
\affiliation[1]{Theoretical Physics Department, CERN, 1 Esplanade des Particules, Geneva 23, CH-1211, Switzerland}
\affiliation[2]{L'Ecole polytechnique f\'ed\'erale de Lausanne, Route Cantonale, 1015 Lausanne, Switzerland}
\emailAdd{kyrylo.bondarenko@cern.ch}

\author[3]{Alexey~Boyarsky,}
\affiliation[3]{Intituut-Lorentz, Leiden University, Niels Bohrweg 2, 2333 CA Leiden, The Netherlands}
\emailAdd{boyarsky@lorentz.leidenuniv.nl}

\author[4]{Marco Nikolic,}
\emailAdd{marco.nikolic@oeaw.ac.at}

\author[4]{Josef~Pradler,}
\affiliation[4]{Institute of High Energy Physics, Austrian Academy of Sciences, Nikolsdorfergasse 18, 1050 Vienna, Austria}
\emailAdd{josef.pradler@oeaw.ac.at}

\author[4]{and Anastasia~Sokolenko}
\emailAdd{anastasia.sokolenko@oeaw.ac.at}

\begin{document}

\abstract{
When the Dark Matter mass is below the eV-scale, its cosmological occupation number exceeds the ones of photons from the cosmic microwave background as well as of relic neutrinos. If such Dark Matter decays to pairs of neutrinos, it implies that experiments that seek the detection of the cosmic neutrino background may as well be sensitive to this additional form of ``dark radiation''. Here we study the prospects for detection taking into account various options for the forecasted performance of the future PTOLEMY experiment. From a detailed profile likelihood analysis we find that Dark Matter decays with lifetime as large as $10^4$~Gyr or a sub-\% Dark Matter fraction decaying today can be discovered. The prospects are facilitated by the distinct spectral event shape that is introduced from galactic and cosmological neutrino dark radiation fluxes. In the process we also clarify the importance of Pauli-blocking in the Dark Matter decay. The scenarios presented in this work can be considered  early physics targets in the development of these instruments with relaxed demands on performance and  energy resolution. 
}

\maketitle

\section{Introduction}

Besides the cosmic microwave background (CMB), the prediction of the cosmic neutrino background (\CNB) is the second, unequivocal key signature of a hot Big Bang.  The Universe must have passed through a stage of billions degrees of Kelvin in order to enable the fusion of light elements from protons and neutrons. At this temperature neutrinos become the main actors balancing the relative abundance of nucleons. However, whereas the measurements of the CMB have elevated Big Bang cosmology to a precision science, the relic radiation of neutrinos from the nucleosynthesis era remains undetected to date. 

The observation of relic neutrinos would provide a window into the first second after the Big Bang, and its detection is  an important task in cosmology. Today's observed CMB radiation temperature of $T=2.73~$K implies that there are 411 relic photons per cm$^3$. Assuming a standard cosmological history, it then follows that the average cosmic \CNB\ number density is 336 cm$^{-3}$ (see e.g.~\cite{Gorbunov:2011zz}).
Relic neutrinos hence constitute the largest of neutrino fluxes at Earth~\cite{Vitagliano:2019yzm}.
However, whereas microwave photons are readily detected,  ``infrared'' neutrinos  have exceedingly small cross section, making them literally inert under any ordinary laboratory scheme. Among the various other ideas for detection~\cite{Stodolsky:1974aq,Shvartsman:1982sn,
Langacker:1982ih,Duda:2001hd,Domcke:2017aqj}, the most prospective way appears to leverage the energy release in the threshold-free electron-neutrino capture reactions of beta-decaying nuclei~\cite{Irvine:1983nr,Cocco:2007za}. Here, the capture on  tritium atoms,
\begin{equation}
\label{capture}
    \nu_e + \text{T} \to \, ^3\text{He} + e^-.
\end{equation}
is one of the best candidate reactions. The relatively low $Q$-value of the associated super-allowed beta-decay, $Q_\beta = 18.529(2)~$keV, allows for schemes that achieve the required sub-eV energy resolution in electron energy, while its half-life $t_{1/2}=12.32(2)~$yrs implies a sensible experimental timescale where sufficient statistics can be collected.

 Because of the minute cross section for neutrino absorption, $ \sigma v_\nu \sim 10^{-44}\,\cm^2$,  detecting relic neutrinos takes extremely large amounts of tritium. The PTOLEMY experiment~\cite{Betts:2013uya,Baracchini:2018wwj} proposes to use 100~g of tritium, which is $10^6$ times more than the best current experiment KATRIN~\cite{Osipowicz:2001sq} employs. Even with this amount of tritium, the expected detection rate is $\sim 4$ or $\sim 8$ relic neutrinos per year, with a dependence if neutrinos are Dirac or Majorana particles~\cite{Duda:2001hd,Long:2014zva}. Both experiments use or plan to use sophisticated schemes that filter electron energies with (sub-)eV precision.  Neglecting the recoil of the daughter nucleus, the electron kinetic energy in reaction~\eqref{capture} is given by
\begin{equation}
    E_e = Q + E_{\nu}.
    \label{eq:EeEnuconnection}
\end{equation}
Any experiment that aims at detecting the neutrino capture~\eqref{capture} in the electron over the common beta background $\rm T \to \, ^3\text{He} + e^- + \bar \nu_e$ must hence resolve an amount $E_\nu$ at the beta-endpoint energy.
For the \CNB---which is guaranteed to be partially composed of massive, non-relativistic neutrinos today---this implies that a successful detection of relic neutrinos is tantamount to measuring neutrino mass. 

Given the long road ahead that the efforts in C$\nu$B detection face, it is only just to ask what other kind of physics can be probed with such experiment. For example, it has been proposed that neutrino capture experiments such as PTOLEMY can be used to detect light sterile neutrinos~\cite{Li:2010sn}, constrain the neutrino lifetime, lepton-asymmetry or thermal history~\cite{Long:2014zva}, or act as directional Dark Matter (DM) direct detectors~\cite{Hochberg:2016ntt,Baracchini:2018wwj}. In this work, we consider yet another possibility, namely, the detection of neutrino ``dark radiation'' (DR). Here, the potentially largest source can be the decay of a fraction $\kappa_\DM$ of DM of mass $m_\DM$.%
\footnote{The possibility of more energetic neutrino DR and its signature in DM direct detection and neutrino experiments was considered in~\cite{PalomaresRuiz:2007ry,PalomaresRuiz:2007eu,Garcia-Cely:2017oco,Cui:2017ytb,Nikolic:2020fom}.} To see, if this is prospective at all, we may saturate the cosmological DR flux by assuming the fraction $\kappa_\DM$ has already decayed (or is currently decaying at an unsuppressed rate) into $N_\nu$ neutrinos---typically $N_\nu = 1,\, 2$ in simple models---to estimate the ratio of DR to \CNB\ neutrino absorption,
\begin{align}
\label{Restimate}
    \frac{R_\DR}{R_\CNB} \sim \frac{N_\nu \kappa_\DM \Omega_{\DM}\rho_{\rm crit}}{m_\DM n_\CNB} \frac{(\sigma c)}{(\sigma v_\nu)}  \sim \mathcal{O}(10) \kappa_\DM \left( \frac{1\,\eV}{m_\DM} \right). 
\end{align}
The first factor is the ratio of the cosmological DM number density over the number density $n_\CNB$ of the \CNB. The second factor is the ratio of absorption cross section times the typical velocity of the incoming neutrino; the latter product is to good approximation velocity independent and hence a number close to unity. This rough estimate neglects the flavor and helicity composition of the \CNB\ as well as additional contributions that may arise from local DM decays%
\footnote{The {\em total} DR flux is independent of DM lifetime $\tau_\DM$ for $\tau_\DM \lesssim t_0$, where $t_0$ is the age of the Universe, and for as long as DR remains relativistic~\cite{Nikolic:2020fom}. 
}%
, but already demonstrates that if DM with mass below the eV-scale decays, it may be detected in a \CNB\ experiment~\cite{McKeen:2018xyz,Chacko:2018uke}. 

The mass-scale of the decaying DM in~\eqref{Restimate} implies that the only channels of decay into Standard Model (SM) particles are photons and neutrinos.  Moreover, when $\kappa_\DM=\mathcal{O}(1)$ the estimate~\eqref{Restimate} implies that we are to consider bosonic light DM. Fermionic  DM needs to satisfy the Tremaine-Gunn bound $m_{\DM} \gtrsim 300$~eV~\cite{Tremaine:1979we,Boyarsky:2008ju} and sub-eV fermionic DM cannot comprise its dominant component.%
\footnote{If $\kappa_\DM\ll 1$ the decaying DM can be fermionic for as long it is only light enough to boost its occupation number by $1/\kappa_\DM$ to retain an $\mathcal{O}(1)$ number in the estimate~\eqref{Restimate}. 
} There are then several possibilities for a concrete realization of this scenario.
A particularly well motivated one is that of a Majoron DM candidate~\cite{Chikashige:1980ui}, for which the decay $\DM \to \nu \nu (\bar \nu \bar \nu)$ is a defining feature. The decay rate is linear in DM mass, $ \Gamma_{\DM\to \nu\nu} \propto g^2 m_\DM $, 
and the required smallness of the effective parameter $g\sim  10^{-17} {\rm eV}/m_\DM$ to achieve a cosmological lifetime arises from the global breaking of lepton number at some UV scale $\langle \Phi\rangle$ that provides neutrino with mass, $g\sim m_\nu/\langle \Phi\rangle$. This possibility was considered in~\cite{McKeen:2018xyz,Chacko:2018uke}. 

In this work, we build on these previous proposals and, first, study  the prospects of an enhanced signature when considering the non-relativistic injection regime, and, second, present a detailed sensitivity study that takes into account the projected performance of PTOLEMY. A Majoron is of course not the only possibility to source low-energy neutrinos. One may equally well consider vector DM particle, associated with a gauged combination of lepton-number and/or baryon minus lepton number, or an eV-scale sterile neutrino with $\kappa_\DM<1$ and enhanced decays to three Standard Model neutrinos through mixing. 
In this work, we will not go into these various options, but rather choose a phenomenological approach, studying the concrete detectability of decaying sub-eV DM; an exploration of models is left for future work.

The  paper is organized as follows: in Sec.~\ref{sec:rate} we introduce the neutrino DR signal in the neutrino capture reactions with a focus on PTOLEMY. In Sec.~\ref{sec:dmdecay} we predict the neutrino DR signal for PTOLEMY, considering various cases. In Sec.~\ref{sec:sensitivity} we forecast the sensitivity of PTOLEMY to detect DR from DM decay, taking into account the backgrounds from beta-decay and relic neutrinos. In Sec.~\ref{sec:conclusions} we summarize our results and conclude.

\section{PTOLEMY neutrino detection rate}
\label{sec:rate}

The overall capture rate (per tritium atom) is given by a product of neutrino flux $n_\nu v_\nu$ times the capture cross section $\sigma$. Importantly, for as long as the incoming neutrino energy satisfies $ E_{\nu} \ll Q$, the product  $\sigma v_{\nu} $ is a constant, with its only dependence on the helicity composition of the incoming flux~\cite{Long:2014zva},
\begin{align}
    n_\nu \sigma v_\nu = \left[ (1-v_\nu) n_{\nu_{h_R}} + (1+v{_\nu}) n_{\nu_{h_L}} \right] (\sigma v)_0 ,
    \label{eq:nsigmav}
\end{align}
with  $(\sigma v)_0 \approx 3.7 \cdot 10^{-45} \text{ cm}^2$ and where $ n_{\nu_{h_L}} (n_{\nu_{h_R}} )$ denote the number density left-helical (right-helical) active neutrinos. This dependence leads to a twice larger event rate when \CNB\  neutrinos are Majorana rather than Dirac. 
For simplicity, in the following we assume that DM decays into equal amounts of right- and left-helical states, which is indeed the case for Majoron decay~\cite{Chacko:2018uke}. By making this assumption, the dependence on helicity composition drops out, simplifying the discussion.%

Neutrinos propagate as mass-eigenstates $\nu_i$ and enter as such the detector in an incoherent mixture of flavor states as they have traveled astronomical distances from the source. The reaction to consider is then $\nu_i + \rm T \to \, ^3{\rm He}+ e^-$ and the probability of capture is  modulated by their  electron-flavor content, given by the squared  PMNS-matrix element $|U_{ei}|^2$. The total rate of neutrino capture in an experiment with a mass $M_T$ of tritium (PTOLEMY plans to use $M_{\text{T}} = 100$~g~\cite{Betts:2013uya}) is given by
\begin{equation}
    \label{eq:GammanuDM}
    \Gamma =  
    \frac{M_{\text{T}}}{m_{\text{T}}}
    \sum_{i=1}^{3}|U_{ei}|^2 \int dE_{\nu,i}\,  \sigma \, v_{\nu,i} \frac{dn_{\nu,i}}{dE_{\nu_i}} 
    \approx 
     \frac{M_{\text{T}}}{m_{\text{T}}}
     (\sigma v)_0 \sum_{i=1}^{3}|U_{ei}|^2 n_{\nu,i}.
\end{equation}
Here, $m_{\text{T}}$ is the mass of one tritium atom, ${dn_{\nu,i}}/{dE_{\nu_i}}$ is the energy spectrum of the $i$-th mass eigenstate; $dE_{\nu_i}$ is the total neutrino energy and $v_{\nu,i}$ the associated velocity. 
For the remainder of this paper, we shall always consider the kinematic regime of low-energy neutrinos where $\sigma_c v_{\nu,i}$ is a constant, and the local number density $n_{\nu,i}$ alone becomes the figure of merit that informs us about the overall rate. 
Applied to the \CNB, the number density of relic neutrinos can be written as $n_{\nu,i} = f_{c,i} n_0$, where $n_0 \approx 56\text{ cm}^{-3}$ is the average number density per neutrino state today, and $f_{c,i}$ is a clustering factor in our Galaxy that ranges from $1$ to $1.1$ for neutrino masses below $50$~meV~\cite{Ringwald:2004np,Akita:2020jbo}.%
\footnote{The absolute neutrino mass scale is unknown; the  best current limit on the sum of neutrino masses is from cosmology, $\sum_i m_{\nu_i}\leq 0.12\,\eV$~\cite{Aghanim:2018eyx}. Therefore, using $50$~meV in the estimate of the clustering factor is already close to the limit for an inverted hierarchy, and clustering is almost negligible.}
The detection rate of  relic neutrinos by PTOLEMY is therefore (c.f.~\cite{Betti:2019ouf})
\begin{equation}
    \Gamma_{\text{CNB}} \approx ( 4\, \text{or}\, 8) \text{ yr}^{-1} \left(\frac{M_{\text{T}}}{100\text{ g}} \right). %
    \label{eq:CNB_rate}
\end{equation}
Here we have taken $ f_{c,i} = 1$ and used unitarity of the PMNS matrix, $ 1= \sum_{i=1}^{3}|U_{ei}|^2$.

The energy spectrum of electrons emitted in the capture is immediately obtained from~\eqref{eq:GammanuDM} together with~\eqref{eq:EeEnuconnection},
\begin{equation}
    \frac{d \Gamma}{d E_e} (E_e) = \frac{M_{\text{T}}}{m_{\text{T}}} (\sigma v)_0  \sum_{i=1}^{3}  |U_{ei}|^2  \frac{dn_{\nu,i}}{dE_{\nu,i}}(E_e-Q).
    \label{eq:DM_rate}
\end{equation}
The detectability of a  signal depends on its intrinsic shape, the values of neutrino masses and on the energy resolution $\Delta$ of the  experiment. We follow~\cite{Betti:2019ouf} and model the latter by a Gaussian with full-width-at-half-maximum (FWHM) given by $\Delta$ to obtain the observed rate,
\begin{equation}
    \frac{d\widetilde{\Gamma}}{dE_{e}}(E_e) = \frac{1}{\sqrt{2\pi} (\Delta/\sqrt{8 \ln 2})} \int \frac{d \Gamma}{dE_{e}}(E') \exp \left( - \frac{(E'-E_e)^2}{2 (\Delta/\sqrt{8 \ln 2})^2} \right) dE'.
    \label{eq:smoothing}
\end{equation}
For the PTOLEMY forecasts below, we shall take $\Delta = 10\dots 100$~eV as a representative range covering the optimal to pessimistic range.

\section{Neutrino dark radiation from Dark Matter decay}
\label{sec:dmdecay}

As we have seen in the previous section, when we consider the additional sources of neutrinos with energies $E_\nu\ll Q$, the relative local number densities inform us about the absolute rates. The rate of DR-induced capture events is hence related to the \CNB\ prediction via,
\begin{equation}
    \Gamma_{\text{DR}} = \frac{n^{\text{DR}}_{\nu_e}}{n^{\text{\CNB}}_{\nu_e}}\Gamma_{\text{\CNB}},
\end{equation}
where $n^{\text{\CNB}}_{\nu_e} \equiv n_0 \sum_i |U_{ei}|^2 f_{c,i}$ is an effective number density of electron neutrinos in the \CNB~  and  $n^{\text{DR}}_{\nu_e} \equiv \sum_i |U_{ei}|^2 n^{\text{DR}}_{\nu,i}$ is an effective number density of electron neutrinos in DR originating from DM decay.

There are then two principal components that source the contributions to $n^{\text{DR}}_{\nu_e}$:
\begin{enumerate}
    \item \textit{DM decay at cosmological distances} -- the averaged neutrino background from DM decays outside of our Galaxy and over the course of history;
    \item \textit{DM decay in the Galaxy} -- neutrinos from DM decays from the DM halo of our Galaxy at present.
\end{enumerate}

As is turns out, for $\tau_\DM\gtrsim t_0$, where $t_0$ is the age of the Universe, these contributions  happen to be of almost identical size in total flux. However, the principal difference is that in the former category the neutrinos experience redshifting of their momenta, whereas in the second category, they do not. This has important consequences for the prediction of the event spectra, and in the following we obtain the respective concentration and energy distribution of both DR components.

\subsection{Neutrino DR from cosmological DM decay}

We first consider the cosmological contribution to the local DR number density. Denoting by $\kappa_\DM n_{\text{DM},0} = \kappa_\DM \Omega_\DM \rho_{\rm crit}/m_{\DM}$ the average number density the decaying DM-component would have today in the limit of infinite lifetime, the number density of the $i$th neutrino mass state can be estimated as%
\footnote{This estimate neglects the effects of the Universe's expansion. The exact expression is given in~\cite{Nikolic:2020fom} and it makes a 50\% downward correction to the estimate for $\tau_\DM \gtrsim t_0$.}
\begin{equation}
    n^{\text{cosm}}_{\nu,i} \approx\text{BR}_i \left(1 - e^{-t_{0}/\tau_{\text{DM}}}\right) N_\nu \kappa_\DM n_{\text{DM},0} \sim \text{BR}_i \frac{t_{0}}{\tau_{\text{DM}}} N_\nu \kappa_\DM n_{\text{DM},0}, \qquad \tau_{\text{DM}}\gg t_{0},
\label{eq:cosmological_neutrino_flux}
\end{equation}
where $\text{BR}_i$ is the branching ratio of DM decay into $i$th neutrino mass state; in the last expression we have exposed the scaling in the limit of long lifetime, $\tau_{\text{DM}}\gg t_{0}$.

Of course, gravity alone already constrains the lifetime of DM, e.g.~from CMB physics. Here, the statement is that either 4\% of all of DM could have decayed between recombination and today, or $\kappa_\DM/\tau_\DM< 6.3\times 10^{-3}\,{\rm Gyr}$ for lifetimes larger than the age of the Universe. The latter implies that $\tau_\DM\gtrsim 12 t_0$ if $\kappa_\DM = 1$~\cite{ Poulin:2016nat}. For simplicity, we either use $\tau_{\text{DM}} =10 t_{0}$ with $\kappa_\DM=1$ for the long lifetime regime, or $\kappa_\DM =0.05$ for the short (arbitrary) lifetime regime,  even if it implies that we slightly slip into the disfavoured region.

Using Eq.~\eqref{eq:cosmological_neutrino_flux} we obtain the effective number density of cosmological electron neutrinos as,
\begin{equation}
    n^{\text{cosm}}_{\nu_e} =
    \sum_{i=1}^3 |U_{ei}|^2 n^{\text{cosm}}_{\nu,i}
    \approx 46\text{ cm}^{-3} 
    \xi
    \left(\frac{10t_{0}}{\tau_{\text{DM}}}\right) \left( \frac{1\text{ eV}}{m_{\text{DM}}} \right),\qquad
    \xi = 3 N_\nu \sum_{i=1}^3 |U_{ei}|^2 \text{BR}_i ,
    \label{eq:ncosme}
\end{equation}
where we have again taken the limit of long lifetime. For equal branching ratios, $\text{BR}_i = 1/3$, the factor $\xi=N_\nu$. Substituting this result into Eq.~\eqref{eq:GammanuDM} the total PTOLEMY detection rate for cosmologically sourced neutrino DR reads
\begin{align}
    \Gamma_\DR^{\text{cosm}} \approx 3.2\text{ yr}^{-1} \xi \left(\frac{M_{\text{T}}}{100\text{ g}} \right)\left(\frac{10t_{0}}{\tau_{\text{DM}}}\right) \left( \frac{1\text{ eV}}{m_{\text{DM}}} \right).
    \label{eq:Gammacosm}
\end{align}
We see that for sub-eV DM mass the detection rate of cosmological neutrinos can be higher than for relic neutrinos, see Eq.~\eqref{eq:CNB_rate}. The scenario can hence become another target for PTOLEMY. It is, however, beyond the reach of KATRIN which uses $M_{\text{T}} = 100~\mu\text{g}$.

To obtain the differential event rate in PTOLEMY, we now turn to the energy spectrum of the cosmological DR component. For simplicity, we shall only consider the case of 2-body decay, $\DM\to \nu\bar\nu$ or $\DM\to \nu\nu$. 
With the initial DM kinetic energy being negligible in comparison to its mass, the neutrinos are initially injected with energy $m_\DM/2$. The associated neutrino momentum then redshifts, and the continuous decay of DM during the cosmic history accumulates to the following spectrum~\cite{Cui:2017ytb},
\begin{equation}
    \frac{dn^{\text{cosm}}_{\nu,i}}{dE_{\nu,i}}(E_{\nu,i}) =\frac{N_\nu}{p_{\nu,i} v_{\nu,i}}   \frac{\text{BR}_i \kappa_\DM n_{\text{DM},0} }{H(z_{\text{dec}})\tau_\text{DM}}  e^{-t(z_\text{dec})/\tau_\DM}  , 
    \label{eq:dnu-dE}
\end{equation}
where $z_{\text{dec}}$ is the decay redshift obtained from the redshift of the initial momentum to the momentum at arrival, $p_{\nu,i} = \sqrt{E_{\nu,i}^2 - m_{\nu,i}^2}$,
\begin{equation}
  z_{\text{dec}} = \sqrt{\frac{\frac{m_{\text{DM}}^2}{4} - m_{\nu,i}^2}{p_{\nu,i}^2}} - 1.
    \label{eq:zdec-E}
\end{equation}
 In the above formula, $t(z)$ is the cosmic look-back time evaluated at $z_{\text dec}$; for $\tau_\DM \gtrsim t_0$ the exponential factor can be neglected. The energy differential flux itself is given by multiplying~\eqref{eq:dnu-dE} by the neutrino velocity~$v_{\nu,i}$.

Examples of the signal for different values of neutrino masses and different assumptions about PTOLEMY energy resolutions are shown (together with the galactic component of the signal that is discussed below) in Fig.~\ref{fig:electron-spectrum-NO} as well as Fig.~\ref{fig:electron-spectrum-IO} in App.~\ref{app:IO}.

\subsection{Neutrino DR from galactic DM decay}
Let us now estimate the contribution to neutrino DR from DM decay inside the Galaxy. For such an estimate it is important  to specify the ratio $v_{\nu}/c$ as it defines the resulting neutrino concentration around us. We will discuss below the two principal cases: neutrinos that escape upon injection and neutrinos that are injected with a speed below the escape speed and are hence retained in the Galaxy.

\subsubsection{Escaping neutrinos}
If neutrinos are injected at velocities $v_\nu$ that exceed the escape speed $v_{\rm esc}\simeq 550\,{\rm km}\,{\rm s^{-1}} \simeq 2\times 10^{-3}c$, they will escape the Galaxy in a time  $t_{\text{esc}} \sim r_{\odot}/v_{\nu} \sim 10^{12}\text{ s}\, (c/v_\nu) \ll t_0$, where $r_\odot = 8.3$~kpc is the distance from the center of the Galaxy to the Sun.
A Galactic contribution with $v_\nu \geq v_{\rm esc}$ is only present for $\tau_\DM\gtrsim t_0$. Taking the long-lifetime limit, an order of magnitude estimate for the local neutrino number density is hence,
\begin{equation}
    n_{\nu,i}^{\text{gal}} \approx  \text{BR}_i \frac{t_{\text{esc}}}{\tau_{\text{DM}}} N_\nu \kappa_\DM n_{\text{DM},\odot} \sim 0.4 \left(\frac{c}{v_{\nu}}\right) \left(\frac{\text{BR}_i}{1/3}\right) n_{\nu,i}^{\text{cosm}} \quad (\tau_\DM \gtrsim t_0). 
    \label{eq:nnugal}
\end{equation}
For this estimate we used the local DM density $\rho_{\text{DM},\odot}\sim 0.3\text{ GeV}/\text{cm}^3$ as a representative value. From this estimate we see
that the Galactic DR contribution can be comparable with the cosmological one and it can be even larger if the DM mass is such that galactic neutrinos are only semi-relativistic. The Galactic flux that takes into account the DM density profile then reads, 
\begin{equation}
    n_{\nu,i}^{\text{gal}} =  \text{BR}_i \frac{r_{\odot}}{v_{\nu,i}\tau_{\text{DM}}} N_\nu \kappa_\DM n_{\text{DM},\odot} \langle D \rangle\quad (v_{\nu,i} > v_{\text{esc}}),
    \label{eq:nnugalDfactor}
\end{equation}
where $\langle D \rangle$ is the whole sky average of the line-of-sight integral as seen from Earth over the galactic DM density distribution (see e.g.~\cite{Cirelli:2010xx}),
\begin{equation}
    \langle D \rangle = \frac{1}{4\pi} \int D d\Omega, \qquad 
    D = \frac{1}{r_{\odot} \rho_{\text{DM},\odot}} \int_{\text{l.o.s.}} \rho(r) ds .
\end{equation}

The value of the averaged D-factor normalised in this way describes how much 
our naive Eq.~\eqref{eq:nnugal} underestimates the local concentration of galactic neutrinos. We adopt $\langle D \rangle = 2.19$ obtained from an NFW profile with a mild dependence on other canonical profiles.%
\footnote{
For the NFW profile we use  $r_s = 24.4$~kpc, $\rho_s=0.18$~GeV$/\text{cm}^3$; for an Einasto profile with the core radius $r_c=1$kpc and $\rho_{\text{DM},\odot}\sim 0.4\text{ GeV}/\text{cm}^3$ one obtains $\langle D \rangle = 2.96$. One may of course entertain the possibility of a core and/or spike at the Galactic center~\cite{Gondolo:1999ef,Ullio:2001fb,Dutton:2014xda}, amplifying the Galactic DR contribution; the effect is  milder than for annihilating DM and we will not go into such possibilities here.}
Similarly to the cosmological case, we  may obtain effective number density of electron neutrinos 
\begin{equation}
    n^{\text{gal}}_{\nu_e} =
    \sum_{i=1}^3 |U_{ei}|^2 n^{\text{gal}}_{\nu,i}
    \approx 43\text{ cm}^{-3} \xi \kappa_\DM \left(\frac{c}{v_{\nu}}\right) \left(\frac{10t_{0}}{\tau_{\text{DM}}}\right) \left( \frac{1\text{ eV}}{m_{\text{DM}}} \right) \quad (\tau_\DM \gtrsim t_0),
    \label{eq:ngale}
\end{equation}
where $\xi$ is given by Eq.~\eqref{eq:ncosme}.
Using Eq.~\eqref{eq:GammanuDM} we estimate the PTOLEMY detection rate of galactic DR neutrinos,
\begin{align}
    \Gamma_{\text{gal}} \approx 3.0\text{ yr}^{-1} \xi \kappa_\DM  
    \left(\frac{c}{v_{\nu}}\right)
    \left(\frac{M_{\text{T}}}{100\text{ g}} \right) \left(\frac{10t_{0}}{\tau_{\text{DM}}}\right) \left( \frac{1\text{ eV}}{m_{\text{DM}}} \right).
    \label{eq:Gammagal}
\end{align}
Comparing Eqs.~\eqref{eq:Gammacosm} and \eqref{eq:Gammagal} we see  that
once the D-factor is taken into account the contributions from  Galactic neutrinos and from cosmological neutrinos sourced by DM decay are almost equal in the long-lifetime regime.

We may at this point check whether we stay clear from any suppression factors that arise from Pauli blocking. The maximum occupation number in a Fermi-Dirac gas is attained from $dn_\nu^{\rm max} = g_\nu/(8\pi^3) d^3 \vec p_\nu$ where $g_\nu=2$ are the  active neutrino degrees of freedom of each massive state. In this section, we consider a kinematic situation where every neutrino, once injected, is on a straight trajectory escaping the Galaxy. In the 2-body decay, a small momentum spread $\Delta p_\nu/p_\nu \sim 10^{-3}$ is inherited from the non-relativistic bound motion of the galactic DM. Integration yields $n_{\nu,\rm gal}^{\rm max}\approx g_\nu/(2\pi^2) p_\nu^3 \times (\Delta p_\nu/p_\nu)$. Taking the ratio with the concentration obtained above yields
\begin{align}
    \frac{n_{\nu,\rm gal}^{\rm max}}{n_{\nu, \rm gal}|_{\rm Eq.~\eqref{eq:nnugalDfactor}}} \approx \frac{10^7}{\kappa_\DM} \left( \frac{v_\nu}{c} \right)^4 \left(\frac{\tau_\DM}{10t_0} \right) \left(\frac{m_\DM}{\eV} \right)^4 ,
\end{align}
and where we have taken the same estimate on the escape time $t_{\rm esc} $ as above. There is a steep dependence  on $m_\DM$ and $v_\nu$, and the ratio may drop below unity, signaling that the naive concentration~\eqref{eq:nnugalDfactor} is affected by Pauli blocking issues. We take this effect into account by multiplying the relevant rates by a phase-space suppression factor,
\begin{align}
    {\rm ps} \approx \frac{n_{\nu,\rm gal}^{\rm max}}{n_{\nu, \rm gal}|_{\rm Eq.~\eqref{eq:nnugalDfactor}}} \leq 1 
    \label{eq:ps_factor}
\end{align}
whenever this ratio drops below unity. We note that this kind of blocking is mitigated in a decay with more than 2 final states, as neutrinos then assume a broader distribution in momentum.

Finally, we may find the optimum injection velocity where the concentration saturates $ n_{\nu, \rm gal}^{\text{max}}(v_{\nu,\text{optimal}}) = n_{\nu, \rm gal} (v_{\nu,\text{optimal}})$
; for this, note that $n_{\nu, \rm gal}^{\rm max}\propto v_\nu^2$ whereas $n_{\nu,\rm gal}\propto 1/v_\nu$ on the account of~$t_{\rm esc}$. Therefore, the ``optimal'' velocity is
\begin{equation}
    \frac{v_{\nu,\text{optimal}}}{c} \approx 4\cdot 10^{-3} \kappa_\DM^{1/3} \left(\frac{10t_{0}}{\tau_{\text{DM}}}\right)^{1/3} \left( \frac{1\text{ eV}}{m_{\text{DM}}} \right)^{4/3} . 
\end{equation}
As can be seen, the number can fall below the escape speed. In this case, the estimate is revised; see following section. The maximal galactic event rate for escaping neutrinos reads,
\begin{equation}
    \Gamma_{\text{gal},\max} \approx 750 \text{ yr}^{-1} |U_{e1}|^2 \left(\frac{M_{\text{T}}}{100\text{ g}} \right) \kappa_\DM^{2/3} \left(\frac{10t_{0}}{\tau_{\text{DM}}}\right)^{2/3} \left( \frac{m_{\text{DM}}}{1\text{ eV}} \right)^{1/3}, \text{ for } m_{\text{DM}} \approx 2 m_{\nu,\text{lightest}}.
\end{equation}
This shows that although Pauli-blocking may severely constrain the absolute numbers, in the optimal case the galactic contribution can far exceed the cosmological one.

\begin{figure}[h]
    \centering
    \includegraphics[width=0.48\textwidth]{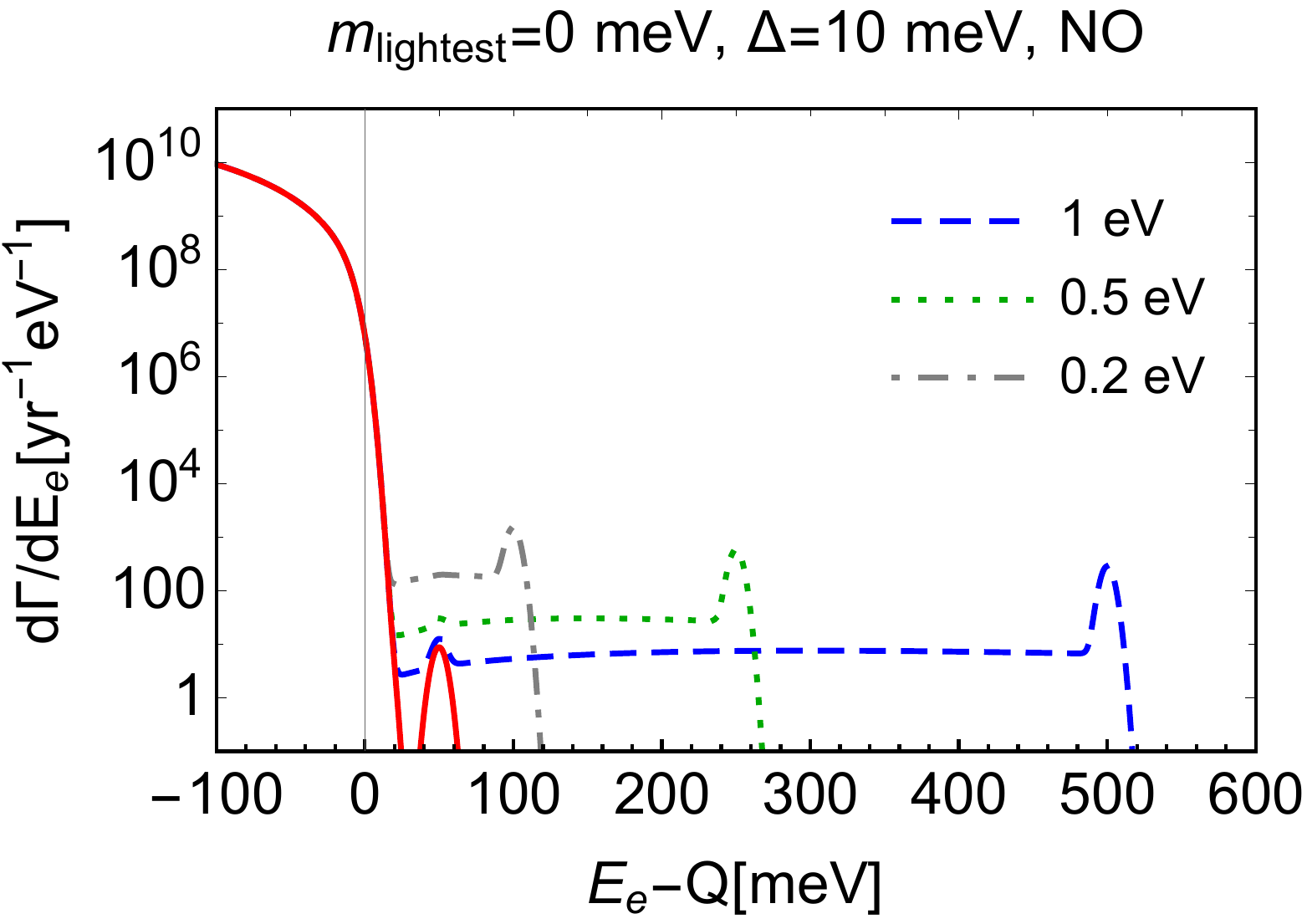}~\includegraphics[width=0.48\textwidth]{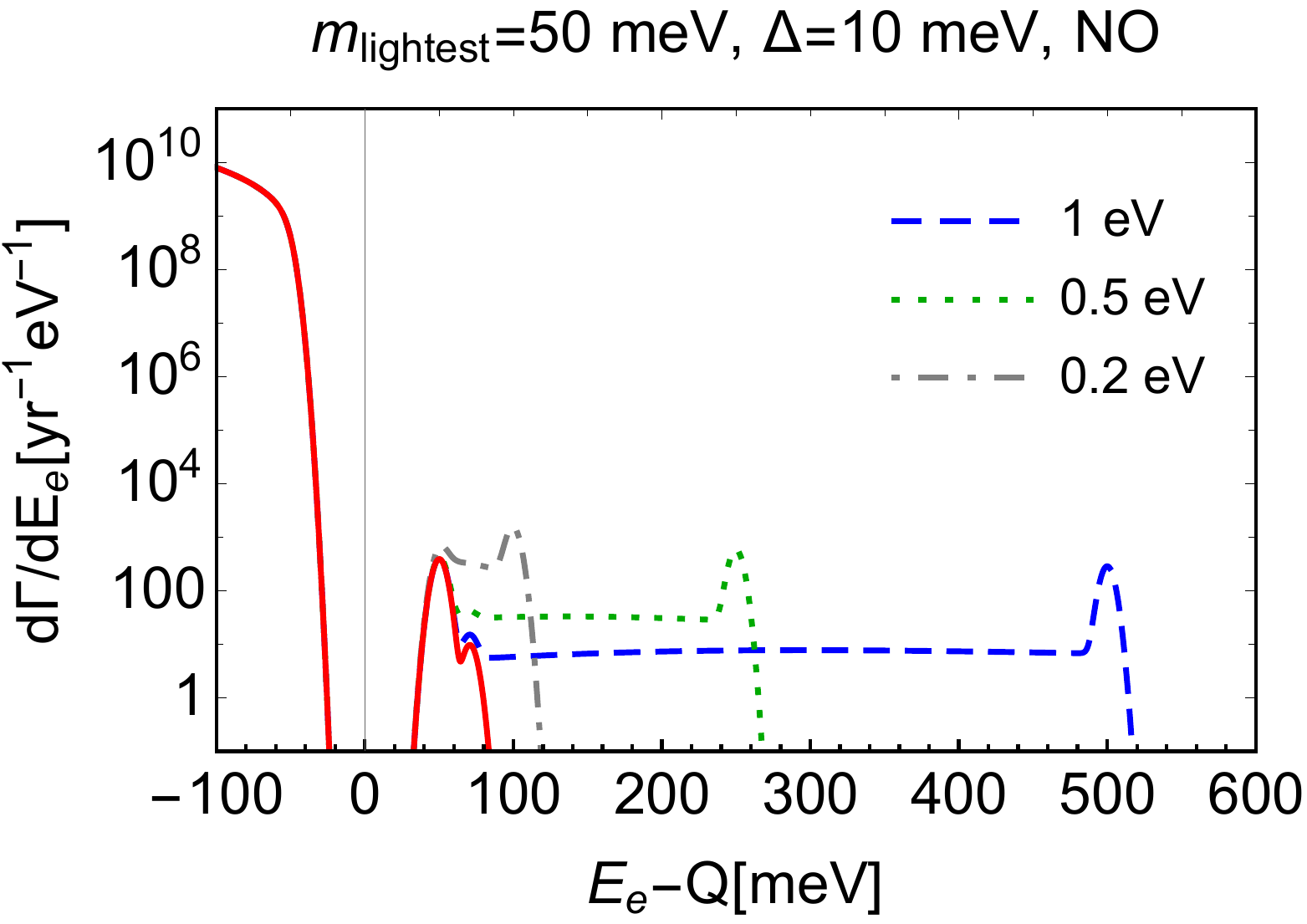}
    \\
    \includegraphics[width=0.48\textwidth]{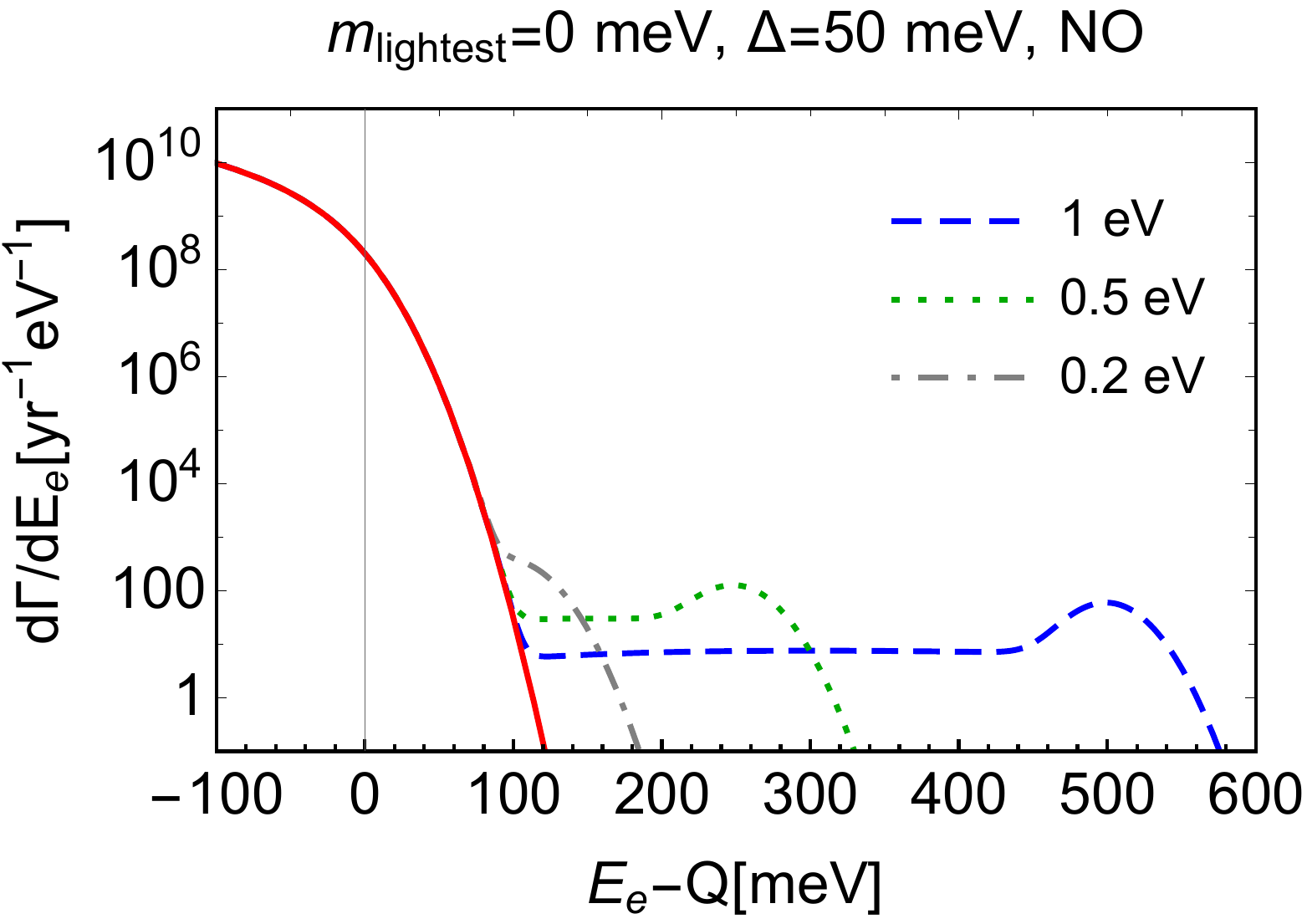}~\includegraphics[width=0.48\textwidth]{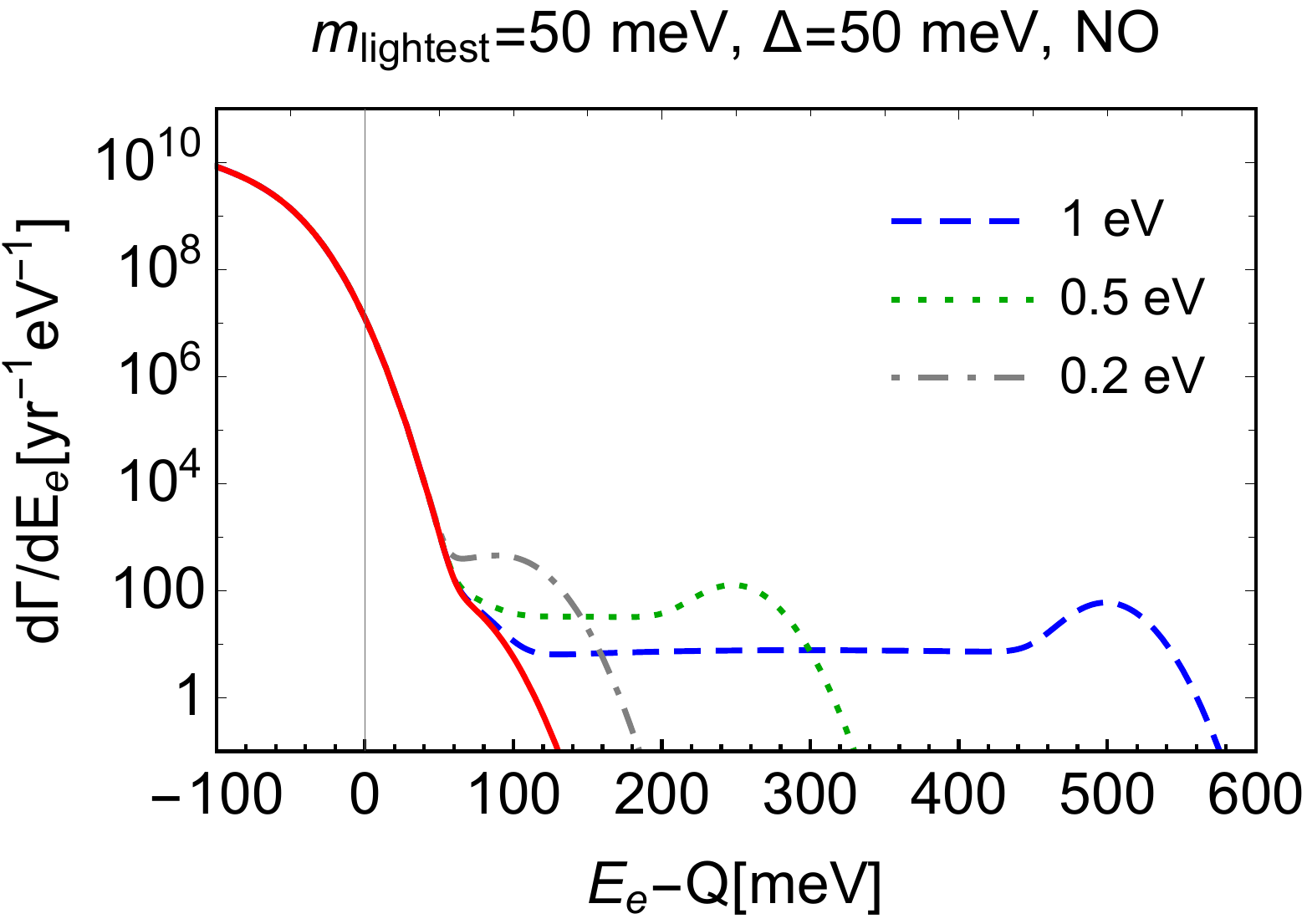}
    \\
    \includegraphics[width=0.48\textwidth]{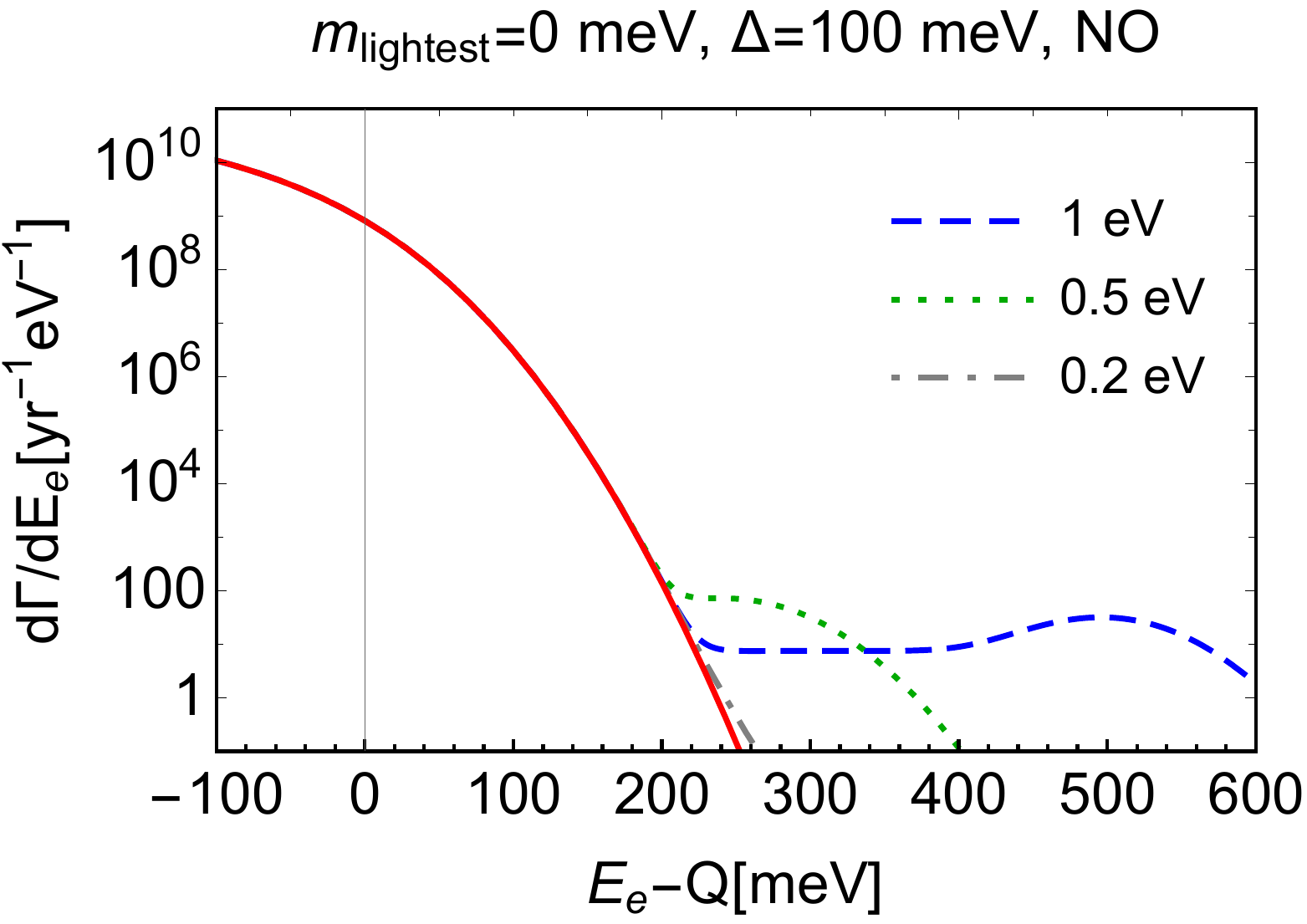}~\includegraphics[width=0.48\textwidth]{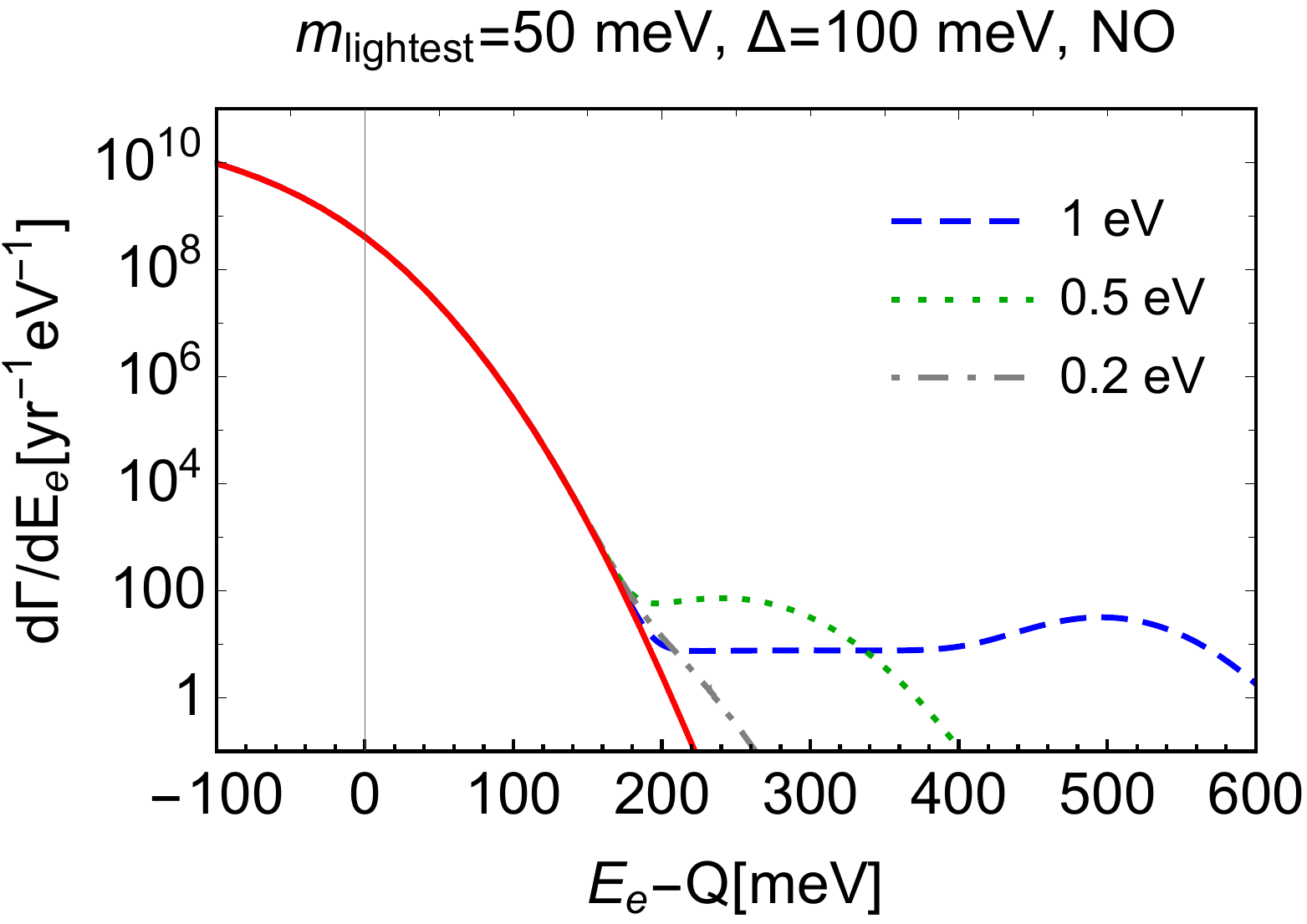}
    \caption{Electron spectra from neutrino DR capture sourced by DM decay with $m_{\text{DM}}=1, 0.5$ and $0.2$~eV and normal neutrino mass ordering (NO) with the mass of the lightest neutrino $m_1=0$ (left panel) and $50$~meV (right panel) and $\xi = 1$. The DM lifetime is taken $\tau_{\text{DM}} = 10 t_{0}$ and the detector energy resolution is varied from $\Delta = 10$~meV, $\Delta = 50$~meV, to $\Delta = 100$~meV from top to bottom.}
    \label{fig:electron-spectrum-NO}
\end{figure}

DR neutrinos from 2-body decay are created with a fixed energy  $E_{\nu}=m_{\text{DM}}/2$ in the DM rest frame.  
In the laboratory frame they will have some energy distribution due to
random velocities of DM particles with the width $\Delta E_{\nu}/E_{\nu} \sim v_{\text{DM}}/c \sim 10^{-3}$.  This scatter in energy  can be neglected compared to the energy resolution of the PTOLEMY detector. Therefore, we may take the Galactic DR neutrino energy distribution as a delta-function,
 \begin{equation}
     \frac{dn^{\text{gal}}_{\nu,i}}{dE_{\nu,i}}(E_{\nu,i}) = n^{\text{gal}}_{\nu,i}\ {\rm ps}\ 
    \delta\left(E_{\nu,i}  - \frac{m_{\text{DM}}}{2}\right).
     \end{equation}
The Galactic component hence  produces a peak at the highest possible energy  for neutrino capture signal,  $E_e=Q + m_{\text{DM}}/2$, see Fig.~\ref{fig:electron-spectrum-NO}.
Assuming an optimistic energy resolution of $\Delta = 10$~meV this peak can be resolved from the cosmological contribution and serves as another signature of the DR scenario. 

\subsubsection{Non-escaping neutrinos}

Let us now consider the special case that neutrinos from DM decay are injected at non-relativistic speeds with
$v_{\nu} \ll v_{\text{esc}}\sim 10^{-3}c$. In this case, neutrinos  accumulate (and saturate in number) rather than escape.
In the 2-body decay benchmark case considered here, it happens when the mass of the lightest neutrino is close to the mass of DM, $m_{\text{DM}}\approx 2 m_{\nu,1}$. 
This kinematic arrangement to yield  $v_\nu\leq v_{\rm esc}$ in the decay, requires fine-tuning,
\begin{equation}
    \frac{m_{\text{DM}}- 2 m_{\nu}}{m_{\text{DM}}} \lesssim 10^{-6} .
    \label{eq:degenerate-m}
\end{equation}
Despite the severeness of the condition, it  may nevertheless be  a natural property in some models 
for the origin of neutrino masses; see e.g.~\cite{Dvali:2013cpa,Dvali:2016uhn}. On the other hand, if DM were to decay into an $n$-body final state with $n>2$, there will always neutrinos with $v_\nu \leq v_{\rm esc}$. Unless the differential decay rate is strongly IR-biased in the neutrino-energy, the efficiency to inject slow neutrinos is directly proportional to the related phase-space volume, which, again will be a small number. Despite these,  at first sight unpalatable circumstances, we shall work out this special case below.

If $v_\nu$ is smaller than the escape velocity from our Galaxy,
$v_{\nu} \ll v_{\text{esc}}\sim 10^{-3}c$, neutrinos rather accumulate than escape. In this case the local neutrino density is\footnote{Gravitationally bound neutrinos can change their helicity when momenta are reversed but spins are not (see e.g.~\cite{Duda:2001hd}). However, as for such neutrinos $v_{\nu} \ll 1$, we can neglect each of the velocity-dependent terms in Eq.~\eqref{eq:nsigmav}.}
\begin{equation}
    n_{\nu}^{\text{gal}} \approx \frac{t_{0}}{\tau_{\text{DM}}} N_\nu \kappa_\DM n_{\text{DM},\odot} \times \text{ps}, %
    \label{eq:n_nu_gal}
\end{equation}
where the escape time is now replaced by the age of the Galaxy, that we have taken as $t_0$ for simplicity. If we neglect the phase space factor ``ps'', Eq.~\eqref{eq:n_nu_gal} suggests a concentration in excess of~\eqref{eq:nnugal} by a factor $t_0/t_{\rm esc}$ which would be enormous.

 Again, when we are sourcing fermionic DR from a bosonic parent with $\kappa_\DM= O(1)$, we need to include the restriction restriction on phase space density by the factor ``ps''. To estimate its importance, let us consider for the parent DM  phase space distribution function a non-truncated Maxwellian, for concreteness again using the local value $n_{\DM, \odot}$,
\begin{align}
    f_\DM (|\vec p|) = \kappa_\DM n_{\DM, \odot} \left( \frac{1}{2\pi m_\DM^2\sigma^2 } \right)^{3/2} \exp\left(-\frac{\vec p^2}{2\sigma^2 m_{\DM}^2}\right).
\end{align}
such that $\int  d^3\vec p \, f_\DM = \kappa_\DM n_{\DM, \odot} $; $\sigma$ is the one-dimensional velocity dispersion. The maximum value is attained for $|\vec p| = 0$ which we denote by $f^{\rm max}_\DM \equiv f_\DM (0) $. In turn, the precise distribution function of the created non-relativistic neutrinos is difficult to know, because the neutrino spends significant time in the Galaxy and is subject to the same thermalization processes as DM. However, if we are to consider a strictly non-relativistic injection with $v_\nu \ll v_{\rm esc}$, which is possible in the finely-tuned 2-body decay, it is not unreasonable to assume that the DM phase space density is largely inherited. Irrespective of the detailed functional form $  f_\nu (|\vec p|)$, however, we may particularly expect that $\tilde f^{\rm max}_\nu \approx  f^{\rm max}_\DM $ holds well, if Pauli-blocking can be neglected.

As mentioned in the previous section, Fermi-Dirac statistics tells us that the maximum phase space density is $f_\nu^{\rm max} = f_\nu/(8\pi^3)$, and  Pauli-blocking in the decay needs to be taken into account, whenever this density becomes saturated. Therefore, we may evaluate the ``in-medium'' phase space suppression factor from the ratio
\begin{align}
    \text{ps} \approx \frac{f_\nu^{\rm max}}{f^{\rm max}_\DM} = \frac{g_\nu}{8\pi^3} \frac{(2\pi \sigma^2)^{3/2} m_\DM^4}{\kappa_\DM\rho_{\DM,\odot}} \approx 4\times 10^{-6} \frac{g_\nu}{\kappa_\DM} \left( \frac{m_\DM}{\rm eV} \right)^4 .
\end{align}
On the right hand side we have used $\sigma = v_c/\sqrt{2}$ where $v_c \simeq 220\,\rm km/\rm s$ is the circular velocity of the solar system. 
This is a punishing factor and implies that the enhancement in the local concentration is at best moderate for $m_\DM\simeq 1\,\eV$, and even turns into a suppression factor for lower DM mass when neutrinos are not evacuated from the galaxy like in the relativistic case above. The arguments above are similar in the spirit that underlie the ones leading to the Gunn-Tremaine bound~\cite{Tremaine:1979we} and essentially a manifestation of Liouville's theorem; see also~\cite{Boyarsky:2008ju}.

Since the solar mass splitting is $\sqrt{|\Delta m_\odot^2|}\simeq 10^{-2}\,\eV$, the degeneracy condition~\eqref{eq:degenerate-m} can only hold for one of the three neutrino mass eigenstates. If the degeneracy holds for the lightest of neutrino states, $\nu_1$, it remains the only kinematically allowed decay channel and $v_\nu \lesssim v_{\rm esc}$ is guaranteed. If the degeneracy is with a heavier state, then the branching ratio into the ``slow channel'' will be suppressed by a model-dependent  factor $\sim (v_\nu/c)^n $, $n\ge 1$ as a lighter final state is available. 
In our numerical results, we will assume for simplicity that the fine-tuning happens for the lightest neutrino in which  case the branching fraction of DM decay into the lightest neutrino mass state $i=1$ is equal to one.

\section{PTOLEMY sensitivity}
\label{sec:sensitivity}

We now proceed to forecast the sensitivity to neutrino DR on the concrete example of PTOLEMY. The canonical event shape for \CNB\ detection is a large beta-background until the endpoint energy that needs to get filtered in order to detect the small capture signals that are offset by a small amount given by their neutrino masses. In the current context, both constitute backgrounds to a DR search. However, as we argued above, the DR signal can extend in energy up to $\sim 1\,\eV$ above the endpoint, into an essentially background free region. 

To treat both cases simultaneously, we use a binned profile likelihood and simulate the experiment with  assumed 1~yr and 5~yr exposures and a target mass of 100~g by generating Monte Carlo mock representations. We consider the neutrino-induced capture events from sources $\alpha = \beta, \CNB$, and $\DR$ together with their associated energy spectra $ d\widetilde \Gamma_\alpha(E_e, \vec \theta)/dE_e $ that are obtained by folding the theoretical rates with an Gaussian energy resolution $\Delta$ according to~\eqref{eq:smoothing}. The model parameters that enter these predictions are $\tau_\DM$ and $m_\DM$ for DR with $\vec\theta=\tau_\DM$; for $\alpha = \beta, \CNB$ $\vec \theta$ there are no fit parameters and $\vec \theta$ is null. The likelihood function under the hypothesis~H for fixed neutrino mass hierarchy (NO, IO), fixed DM mass $m_\DM$ and absolute neutrino mass scale given by $m_{\nu_1}$ reads,
\begin{align}
  \label{eq:LeventsBin}
  \mathcal{L}(\vec \theta | {\rm H}) = \prod_{i=1}^{N_{\rm bin}}  \frac{e^{- \varepsilon  \sum_{\alpha} \mu_{\alpha}^i (\vec \theta)}  }{N_{\rm obs}^i!} \, \bigg[\sum_{\alpha} \mu_{\alpha}^i (\vec \theta) \bigg]^{N_{\rm obs}^i}. 
\end{align}
 For our analysis, we divide the signal region in $E_e-Q$ from $-25, -75,-150$~meV (for $\Delta=10,50,100$~meV, respectively) to $300$~meV into $N_{\rm bin} = 100$ equidistant bins and from $0.3$~eV to $30$~eV into $N_{\rm bin} = 50$ logarithmic bins. The reason for such division is owed to computational efficiency, since at higher energies we enter the background-free region. The expected number of events in each bin~$i$ of source $\alpha$
is denoted by  $\mu_{\alpha}^i $ and $N_\alpha^i$ is the associated  random number of observed events in each bin that is drawn from a Poisson distribution;  $N_{\rm obs}^i= \sum_\alpha N_\alpha^i$.

A discovery of a DR signal in presence of backgrounds then amounts to a rejection of the background-only hypothesis $H_0$ for sources $\alpha= \beta, \CNB$.
Here, the negative log-likelihood then serves as test statistic for the hypothesis test,
$q=-2 \ln { \mathcal{L}( \hat{\hat{\vec \theta}} | {\rm H_0}) }/{ \mathcal{L}( \hat {\vec \theta} | {\rm H_1})},$
where $\hat{\hat{\vec \theta}}$ maximizes the likelihood under the background-only hypothesis H$_0$: $\tau_\DM \to \infty$ and $\hat {\vec \theta}$ maximizes $\mathcal{L}$ for  signal plus background, H$_1$: $\tau_\DM\neq 0$. A distribution in $q$ under H$_1$ is obtained by generating $10^3$ mock data-sets for each combination of $(\tau_\DM,m_\DM)$, until the entire parameter space is scanned; in turn, the distribution in $q$ with mock-data generated under H$_0$ we verified that it follows a $\chi^2$ distribution with one degree of freedom as per Wilk's theorem \cite{wilks1938}. The significance distributions are then given by $Z = \sqrt{q}$. 
The discovery criterion at 3$\sigma$ significance implies that H$_0$ is rejected with $99.865\%$ probability ($p$-value $p_0=0.00135$). For a chosen confidence level of $90\%$ we require that a given experiment has a $90\%$ probability to detect at least a signal with 3$\sigma$ significance. Hence, this leads to a detection of the signal, if $90\%$ of the mock-data sets generated under H$_1$ lie above the discovery criterion $Z \geq 3\sigma$, where H$_1$ is accepted and where at the same time H$_0$ is rejected; see~\cite{Billard:2013qya,Nikolic:2020fom} for further details on this procedure.

\begin{figure}[h]
    \centering
\includegraphics[width=0.5\textwidth]{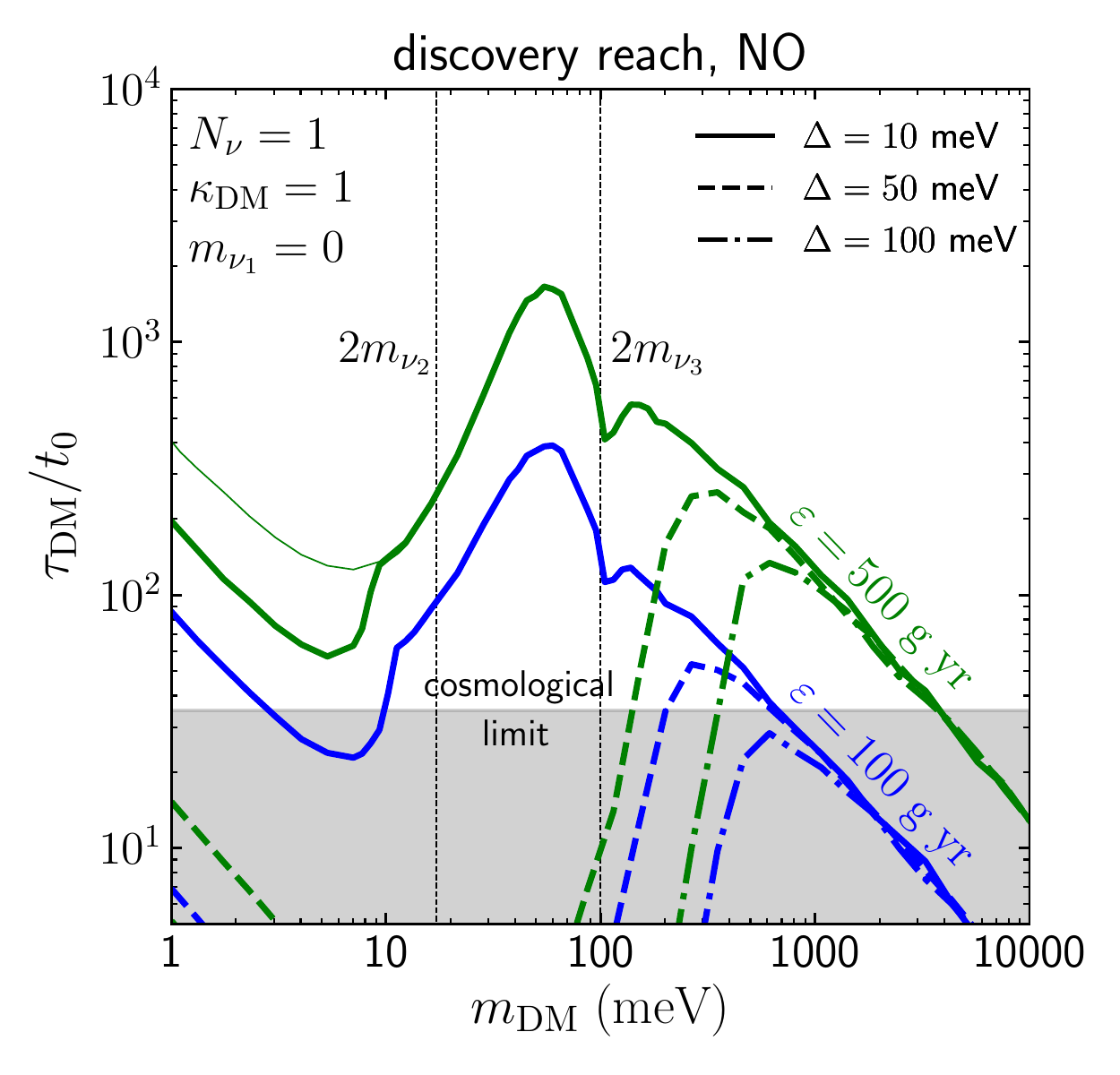}%
\includegraphics[width=0.5\textwidth]{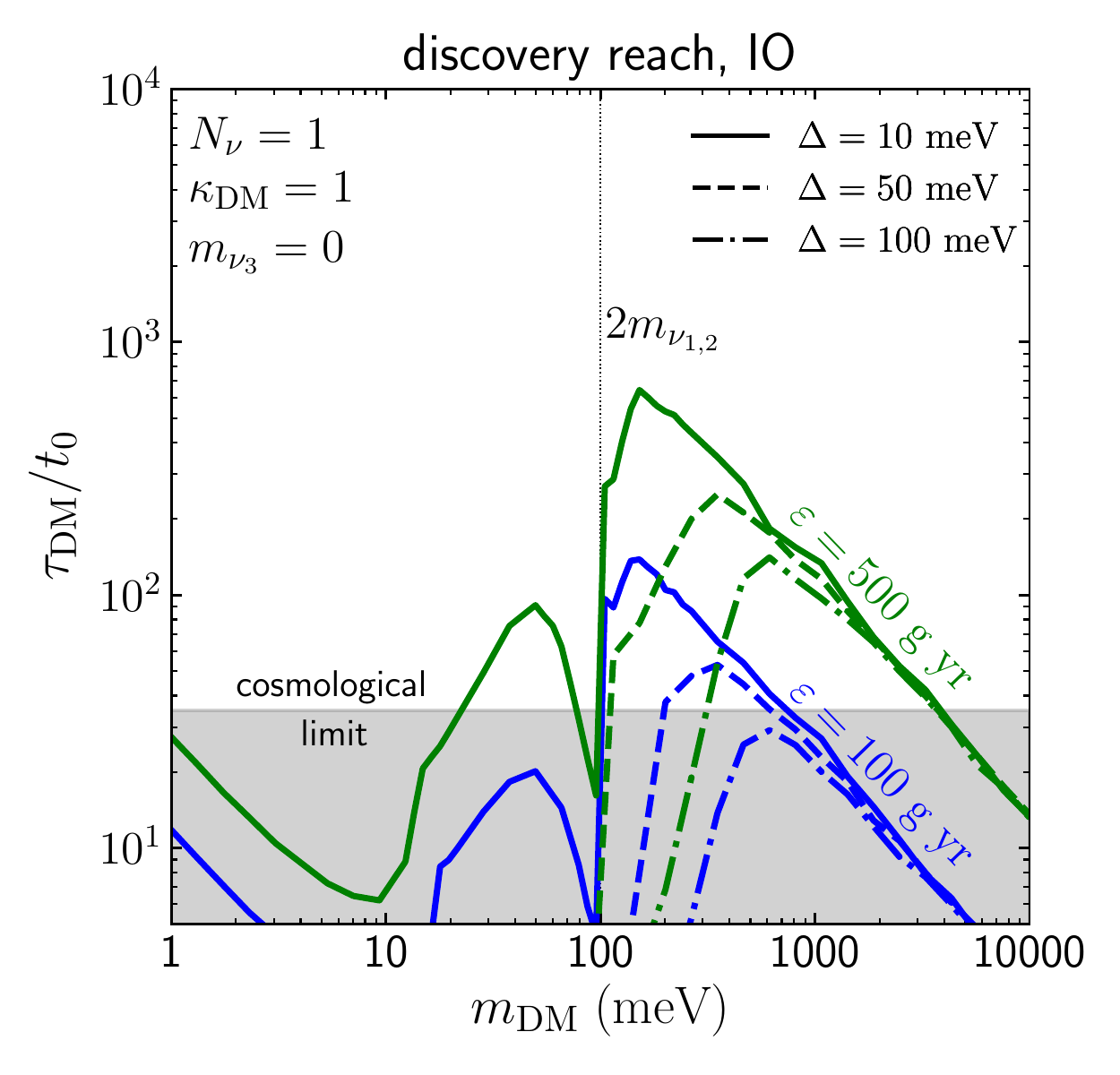}
    \caption{\small Discovery reach of PTOLEMY as a function of progenitor mass $m_{\rm DM}$ and lifetime in units of the age of the Universe, $\tau_{\rm DM}/t_0$. An exposure of 100~g~yr (blue lines) and 500~g~yr (green lines) has been assumed for various projected performances on the electron energy resolution $\Delta$ as labeled, with  10~eV (100~eV) being the optimal (most conservative) case. All of DM is assumed to be decaying $\kappa_{\rm DM} = 1$. The mass of the lightest neutrino is $m_{\nu_1}=0$. In the left (right) panel The mass of the lightest neutrino is $m_{\nu_1}=0$ ($m_{\nu_3}=0$) and a normal (inverted) hierarchy is assumed.} %
    \label{fig:limits}
\end{figure}

\begin{figure}[h]
    \centering
\includegraphics[width=0.5\textwidth]{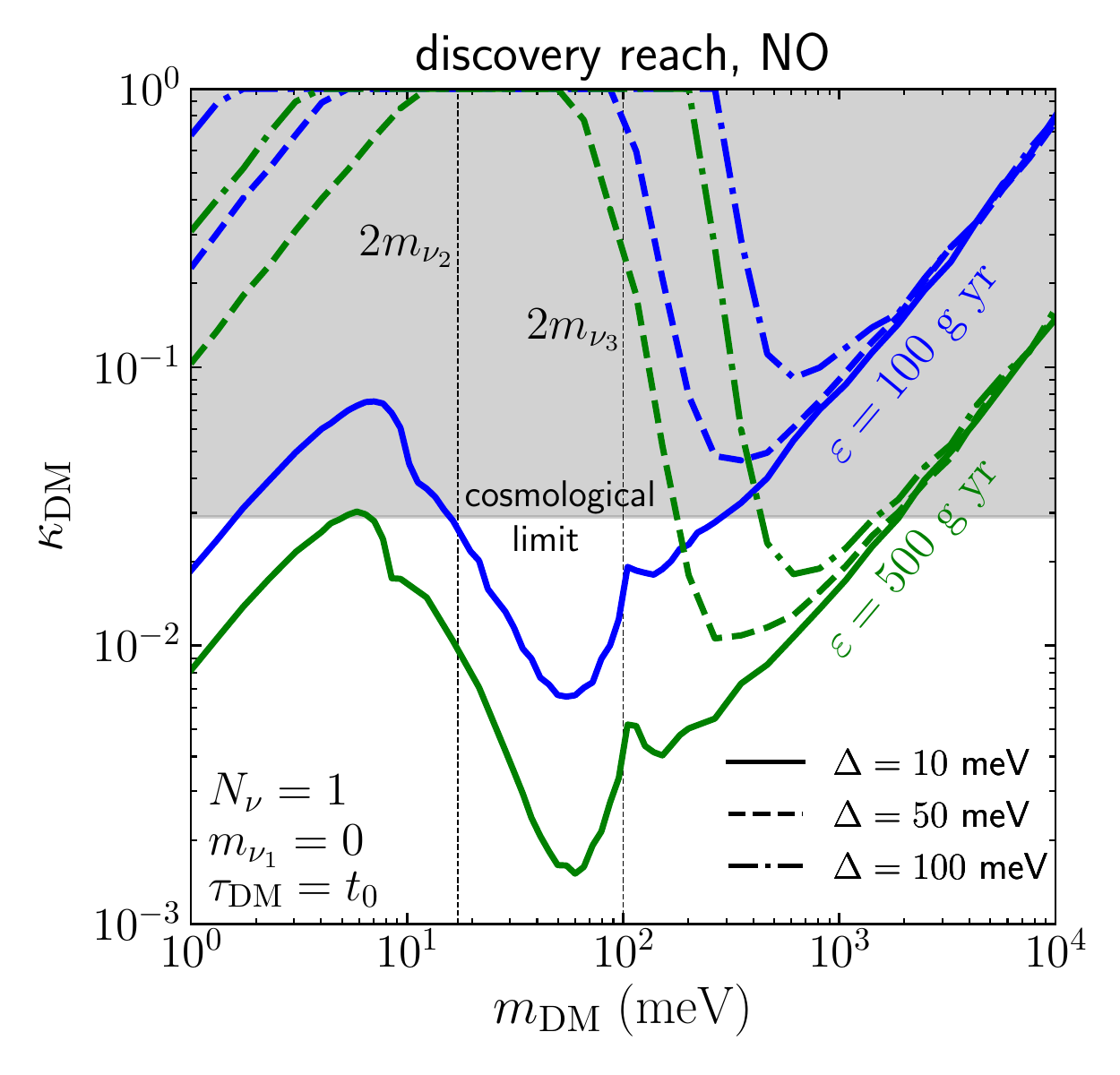}%
\includegraphics[width=0.5\textwidth]{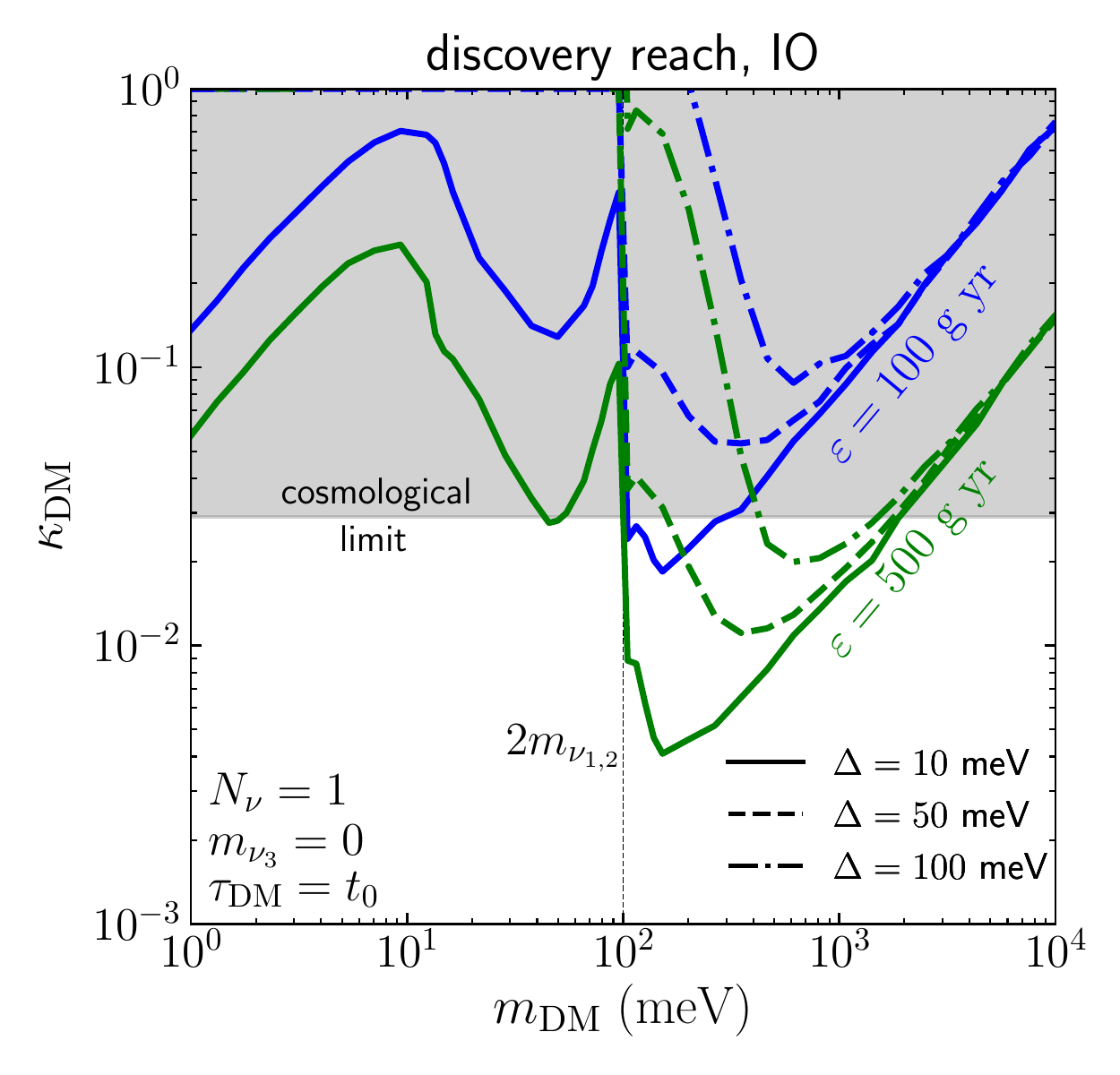}%
    \caption{\small Sensitivity of PTOLEMY to the decaying fraction of DM, $\kappa_{\rm DM}$, as a function of DM mass; same labeling as in Fig.~\ref{fig:limits}.}
    \label{fig:discFrac}
\end{figure}

The resulting discovery potentials are shown in the left (right) panel of Fig.~\ref{fig:limits} for normal (inverted) neutrino mass hierarchy; we additionally take the lightest neutrino as massless, $m_{\nu_1}=0$ ($m_{\nu_3}=0$). We assume that  DR is sourced from $X\to \nu\bar \nu $, i.e.~$N_\nu=1$ with a decaying DM fraction of  $100\%$, $\kappa_{\rm DM} = 1$.
The blue and green sets of lines are associated with exposures of 100~g~yr and 500~g~yr, respectively. The solid, dashed, and dash-dotted lines correspond to progressively worsening assumptions on the energy resolution, $\Delta = 10,\ 50$ and 100~meV, respectively. The thin vertical lines show the kinematic thresholds for the decays into the heavier neutrino mass eigenstates $m_{\nu_{2,3}}$. The gray shaded region shows the  cosmological limit on the   decaying cold DM lifetime, $\tau_\text{DM} \gtrsim 35 t_0$~\cite{Chen:2020iwm}; see also~\cite{Enqvist:2015ara,Poulin:2016nat,Nygaard:2020sow}. Finally, the thin green line in the left panel is obtained when Pauli-blocking is neglected.

Both panels establish the sensitivity to the maximum DM lifetime, directly related to the minimum detectable DR flux (a $3\sigma$ significance in the general presence of the \CNB\ background. In the high mass region  $m_\text{DM} \gtrsim 100~\mathrm{meV}$, the discovery potential is almost independent on the neutrino mass hierarchy and a general $1/m_\text{DM}$ scaling can be seen. At around $100~\mathrm{meV}$ progenitor mass, this trend is broken by the presence of  
the $\CNB$ peak generated by the heaviest neutrino, %
$m_{\nu_3} =50~\mathrm{meV}$ ($m_{\nu_{1,2}} \simeq 50~\mathrm{meV}$) assuming normal (inverted) hierarchy.
In the left panel, the sensitivity for $\Delta=10~\mathrm{meV}$ reaches its optimum at $m_\text{DM} \sim 50~\mathrm{meV}$, whereas in the inverted hierarchy scenario (right panel) the lightest neutrino has a smaller contribution to the DR signal due to the smaller squared PMNS-matrix element $|U_{e3}|^2$. Therefore, smaller lifetimes, i.e.~a larger DR flux, are necessary to discover the signal in comparison to the right panel. For $m_\text{DM} \lesssim 40~\mathrm{meV}$, the continuous tritium beta background starts playing a role, suppressing the lifetime reach in both panels. However, this is eventually counterbalanced by the growing decaying DM number density with $1/m_{\rm DM}$ and the sensitivity is again improved for diminishing DM mass. However, only for the NO the lines extend above the cosmological limit.

We conclude that a discovery of decaying DM with $m_\text{DM} \gtrsim 100~\mathrm{meV}$ is possible with rather relaxed assumptions on energy resolution. For lighter DM mass, an optimum energy resolution is critical to suppress the bleeding of the beta background into the signal region, and decaying DM with neutrino final states is discoverable in the normal ordering across the entire conceivable mass range.

Finally, we may consider the possibility that a fraction $\kappa_{\rm DM}$ of DM decays with arbitrary lifetime and ask for the sensitivity of PTOLEMY to $\kappa_{\rm DM}$. For this, we saturate the flux by choosing an optimal lifetime, $\tau_\text{DM}=t_0$, so that the fraction $\kappa_{\rm DM}$ decays today with an unsuppressed rate.  The model parameters that enter in the likelihood in Eq.~\eqref{eq:LeventsBin} are now $\kappa_\text{DM}$ and $m_\text{DM}$ for DR with $\vec{\theta}=\kappa_\text{DM}$. Figure~\ref{fig:discFrac} presents the $3\sigma$ discovery sensitivity to $\kappa_{\rm DM}$ as a function of progenitor mass as above.  As expected, the discovery potentials in the $(\kappa_\text{DM}, m_\text{DM})$-plane exhibit an inverse behaviour with respect to the contours in the $(\tau_\text{DM}, m_\text{DM})$-plane in Fig.~\ref{fig:limits}. The discoverable region is hence affected by the same limiting factors as were discussed above. We conclude that with an exposure of 100~g~yr (500~g~yr) PTOLEMY is capable to detect a decaying fraction of $\sim 1\%$ ($\sim 0.1\%$) with an optimal energy resolution of $\Delta = 10$~meV. For the pessimistic case $\Delta = 100$~meV it takes the larger of assumed exposures to compete with cosmological limits with a mild prospect to detect DR originating from a decay with progenitor mass $0.1 \lesssim m_{\rm DM}/{\rm eV}\lesssim 2$.

\section{Conclusions}
\label{sec:conclusions}

PTOLEMY is a visionary and ambitious experiment. Its main science goal -- the detection of relic neutrinos -- would mark a resounding success for a key prediction of hot Big Bang cosmology, 
but will require significant breakthroughs in experimental technology.
When entering such unexplored areas we are not safe from unexpected difficulties and obstacles.
In this work we demonstrate through a detailed profile likelihood study that even before PTOLEMY reaches the level of performance 
(first of all, energy resolution and statistics) it can potentially detect a signal from new physics that can accede  the SM relic neutrino signal, namely, the detection of neutrino DR. Such DR may be sourced by the decay of (a component of) DM with sub-eV mass. The potential signal in PTOLEMY can then be classified as follows:
\begin{itemize}
    \item In the most generic case (see Fig.~\ref{fig:electron-spectrum-NO}) DM decays into a 2-body neutrino final state which results in an additional  peak located  at $E_{\text{peak}}=Q+ m_{\text{DM}}/2$. The number of events in this signal may be equal or larger than in the signal from the \CNB~ for the DM mass range $2 m_{\nu_1}\leq m_{\text{DM}}\lesssim 1$ eV. This peak comes from DM decay in our Galaxy. Additionally, extragalactic DM decays give rise to a second component of the signal, similar in overall magnitude but with
    events almost equally distributed in electron energy between $Q$ and $Q+ m_{\text{DM}}/2$. 
    \item There is a special case when neutrinos are efficiently released at semi- or non-relativistic velocities, either in a suitably arranged decay with more than two final states, or when $m_{\text{DM}}/2\simeq m_{\nu_1}$ in the 2-body decay. The local concentration of neutrinos can then be enhanced by Galactic DM decays, reaching a maximum when the injection velocity is in the vicinity of the Milky Way's escape speed. 
\end{itemize}
There is a number of avenues to explore further in our proposal. First of all, concrete models of sub-eV DM should be explored and how they embed themselves into the bigger scheme of things, such as relic density generation; such program has already started in~\cite{McKeen:2018xyz,Chacko:2018uke}.  Is it possible to find well-motivated or natural cases where the non-relativistic injection boosts the detection prospects? On the signal side, we may quantitatively address the question to what degree it is possible to discriminate between early ($\tau_\DM \lesssim t_0$) and late ($\tau_\DM \gtrsim t_0$) decays by virtue of the Galactic peak. In summary, there is a scientific case for relic neutrino searches such as PTOLEMY that is connected to another pressing topic in modern physics, namely, the quest in understanding the most basic properties of DM, such as its lifetime and mass-scale. A detection of DR in a future \CNB\ experiment may shed light on these questions.

\paragraph{Acknowledgements} MN and AS are supported by the FWF Research Group grant FG1. JP is supported through the New Frontiers Program by the Austrian Academy of Sciences. KB and AB are supported by the European Research Council (ERC) Advanced Grant ``NuBSM'' (694896).

\appendix

\section{A solar neutrino basin?}

Another intense local source of neutrinos is the Sun. Here, one may first wonder if the most prominent of fluxes, the $pp$-flux may constitute a background for PTOLEMY. However, it is easy to see that the falling statistical beta spectrum with decreasing energy yields a small flux, e.g.~at $E_\nu = 1\,\eV $ it is $dF_{pp}/dE_\nu \simeq 10^{-2}\,{\rm cm^{-2}\,s^{-1}\,keV^{-1}}$. Overall, the spectrum  translates into a minute local concentration $n_{\nu, \rm pp} \sim 10^{-16} $ of $pp$-neutrinos below 1~eV energy. In fact, the low energy region is largely dominated by the flux from  plasmon decay~\cite{Raffelt:1996wa}. From Fig.~2 in~\cite{Vitagliano:2017odj} one finds a differential flux value $dF/dE = 10 \text{ cm}^{-2} \text{ s}^{-1} \text{ keV}^{-1}$ at a neutrino energy of $1$~eV, still falling significantly short for an interesting instantaneous concentration. 

The above arguments neglect neutrino mass. Neutrinos produced in the Sun may, however, also be gravitationally trapped within the solar system because of their finite  masses. Such possibility has recently been proposed in~\cite{VanTilburg:2020jvl} as an amplification scheme for probing light new physics that may be produced in Sun. We may take a quick estimate to demonstrate, that for neutrinos this mechanism is negligible to obtain a reasonable neutrino concentration at Earth.

Let the flux of neutrinos that reaches Earth but do not escape  the solar system be~$\Delta F$. On dimensional grounds, the number density of neutrinos at Earth is then of order,
\begin{equation}
    \label{eq:nusolar}
    n_{\nu,\text{solar}} \sim \frac{\Delta F t_{\odot}}{r_E},
\end{equation}
where $t_{\odot} \approx 4.5 \times 10^9\text{ yr}$ is the age of the solar system and $r_E=1\,{\rm AU}$ is the distance between  Sun and Earth.
To estimate the neutrino flux we may take the above quoted value of $10 \text{ cm}^{-2} \text{ s}^{-1} \text{ keV}^{-1}$ at 1~eV energy from~\cite{Vitagliano:2017odj}, hence overestimating the relevant non-relativistic portion at even lower energy.%
\footnote{It appears that finite neutrino masses were not taken into account in the numerical results of~\cite{Vitagliano:2017odj}, but their inclusion would render the neutrino flux even smaller.}
Neutrinos that reach Earth but do not escape the solar system have a narrow energy distribution with a width 
\begin{equation}
    \Delta E = \frac{G M_{\odot} m_{\nu}}{r_E} \approx 10^{-8} \text{ eV} \left(\frac{m_\nu}{1\text{ eV}}\right) .
\end{equation}
We may then limit the associated flux of such neutrinos from above,
\begin{equation}
    \Delta F \lesssim \frac{dF}{dE}  (E=1\text{ eV})\Delta E  \approx 10^{-10} \text{ cm}^{-2} \text{ s}^{-1} \left(\frac{m_\nu}{1\text{ eV}}\right)
\end{equation}
Substituting this value into~\eqref{eq:nusolar} we arrive at
\begin{equation}
    n_{\nu,\text{solar}} \sim 10^{-6}\text{ cm}^{-3}.
\end{equation}
This number is already overestimation of the trapped neutrino density at Earth and it is eleven orders of magnitude smaller than for relic neutrinos. This means that solar neutrinos do not constitute a background for relic neutrino searches.

\section{Inverse Ordering}
\label{app:IO}
\begin{figure}[h]
    \centering
    \includegraphics[width=0.48\textwidth]{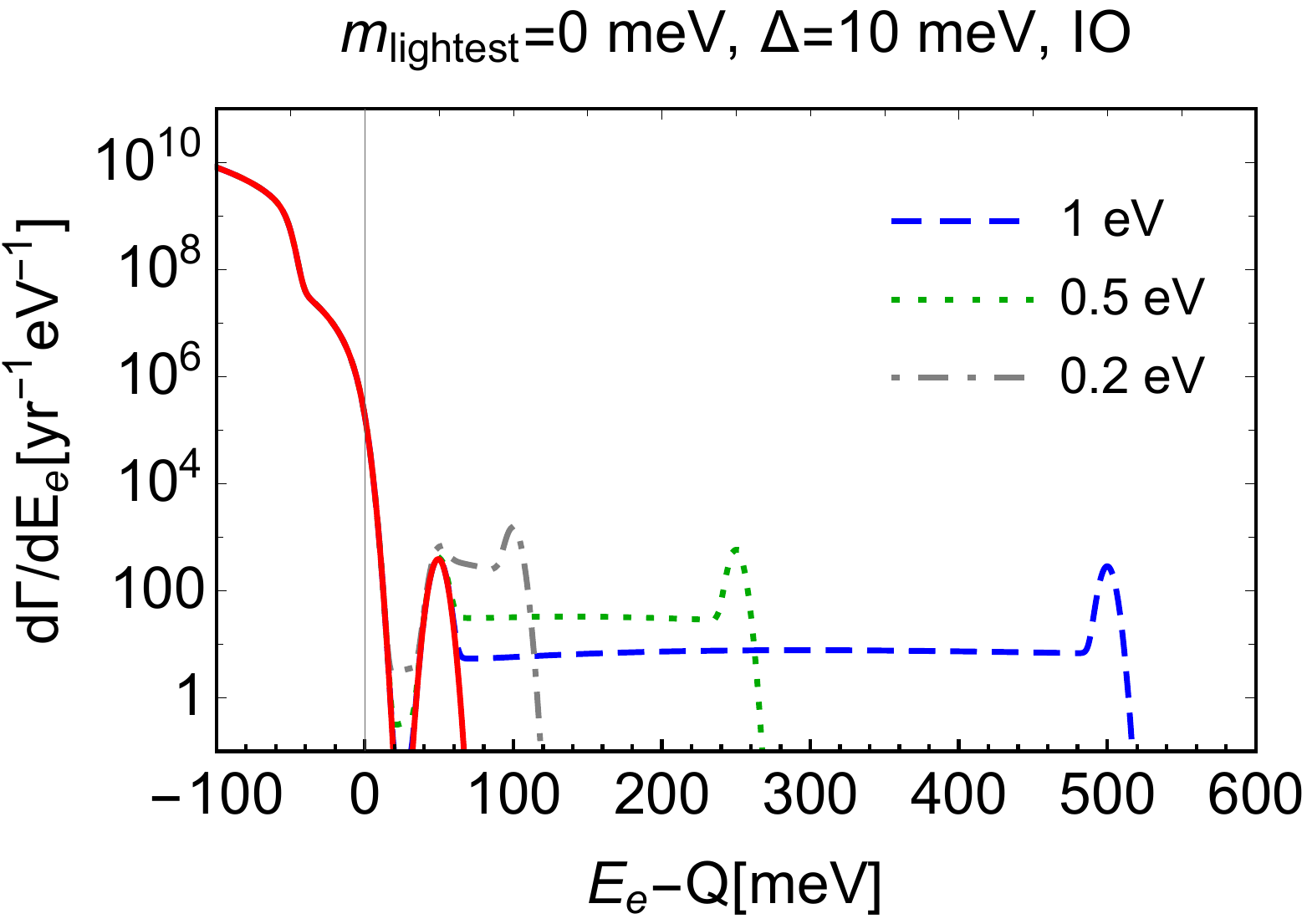}~\includegraphics[width=0.48\textwidth]{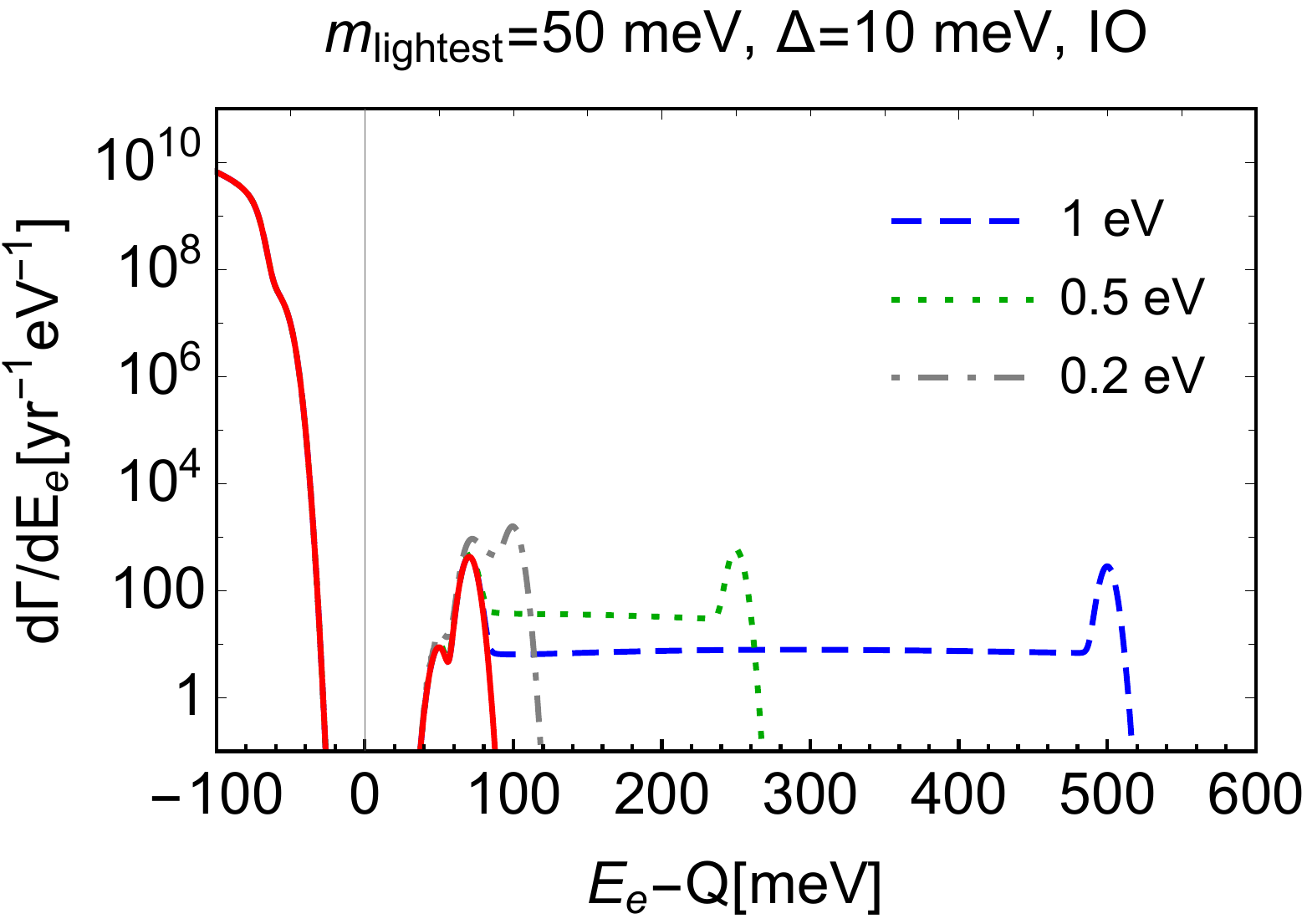}
    \\
    \includegraphics[width=0.48\textwidth]{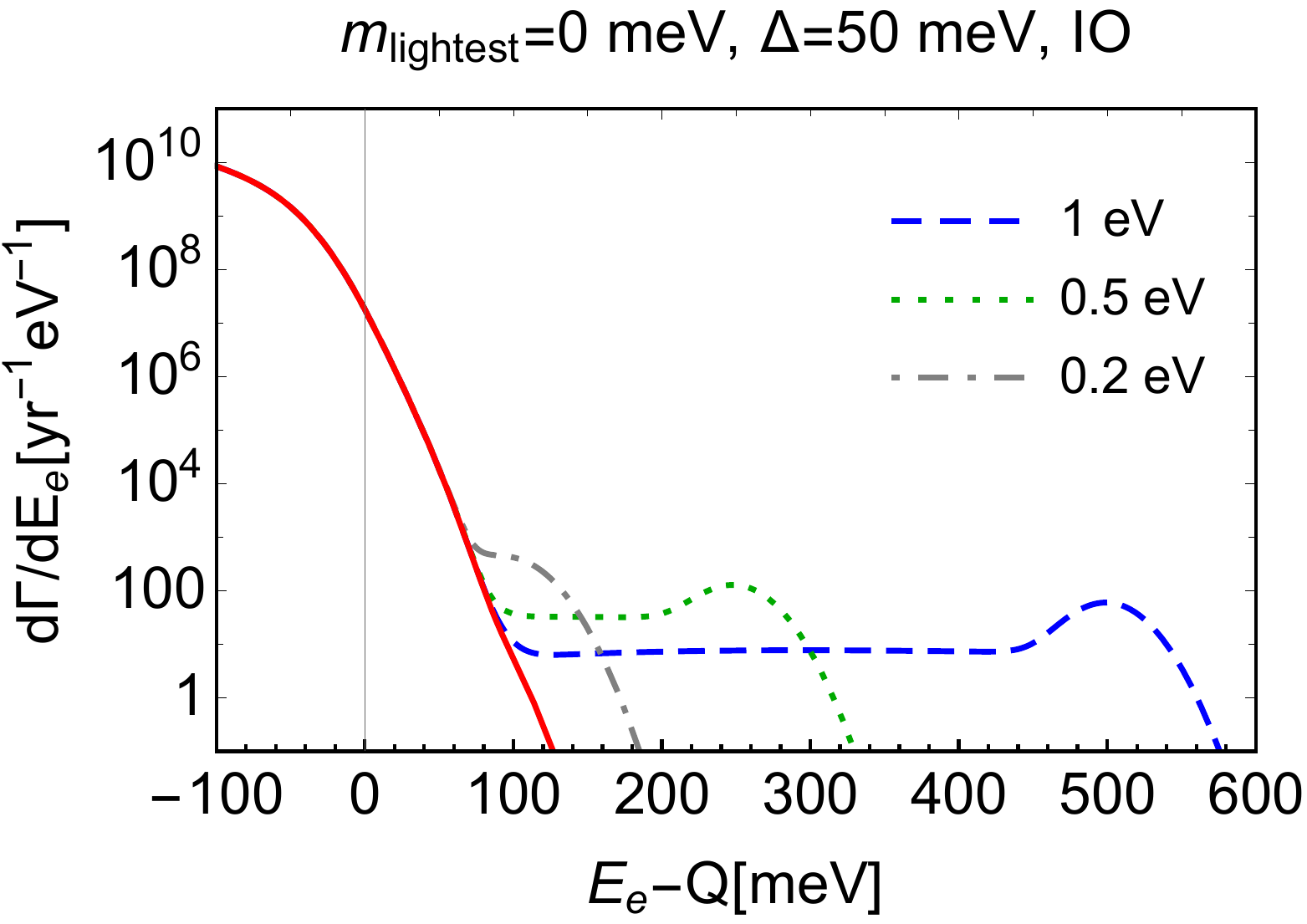}~\includegraphics[width=0.48\textwidth]{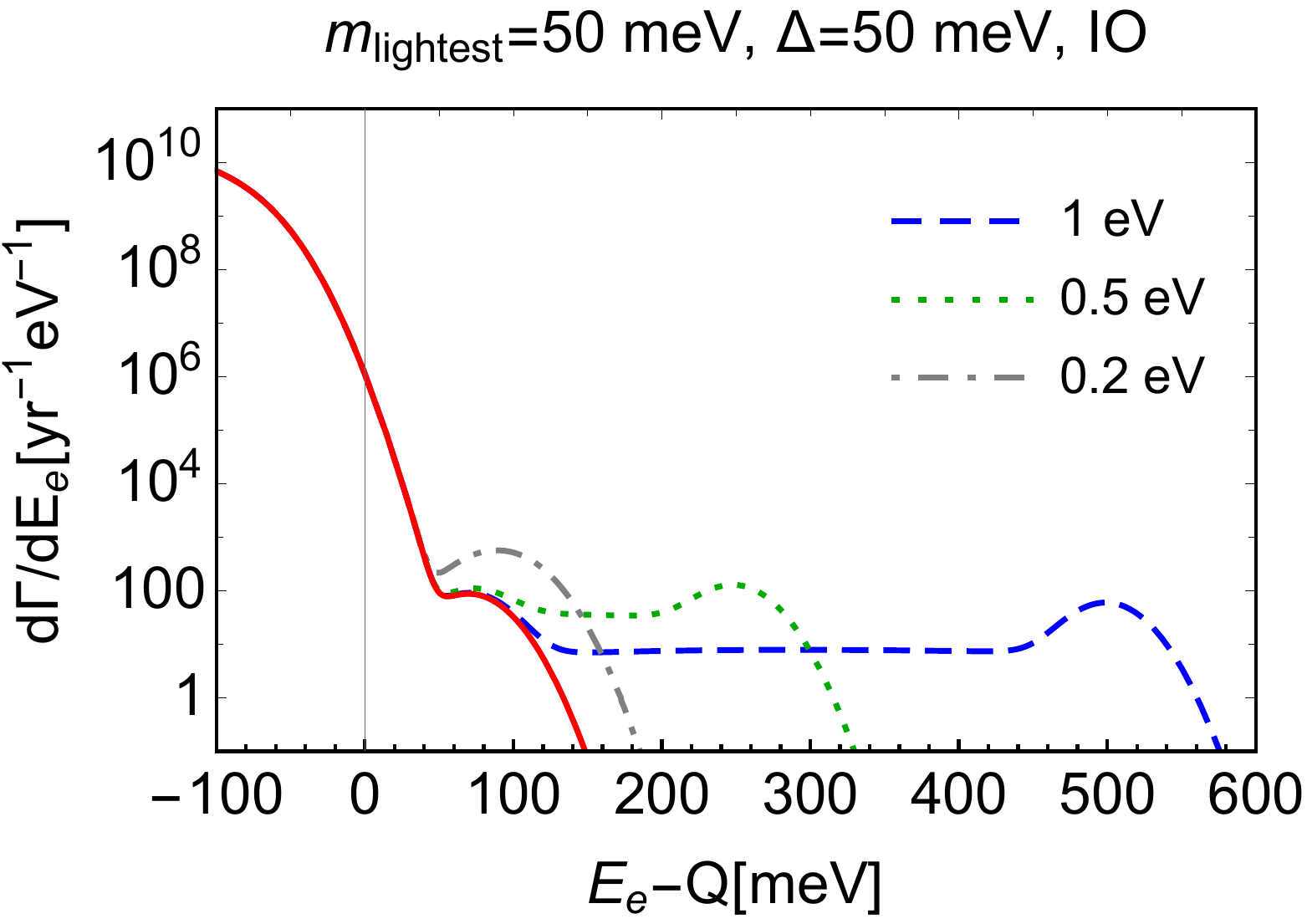}
    \\
    \includegraphics[width=0.48\textwidth]{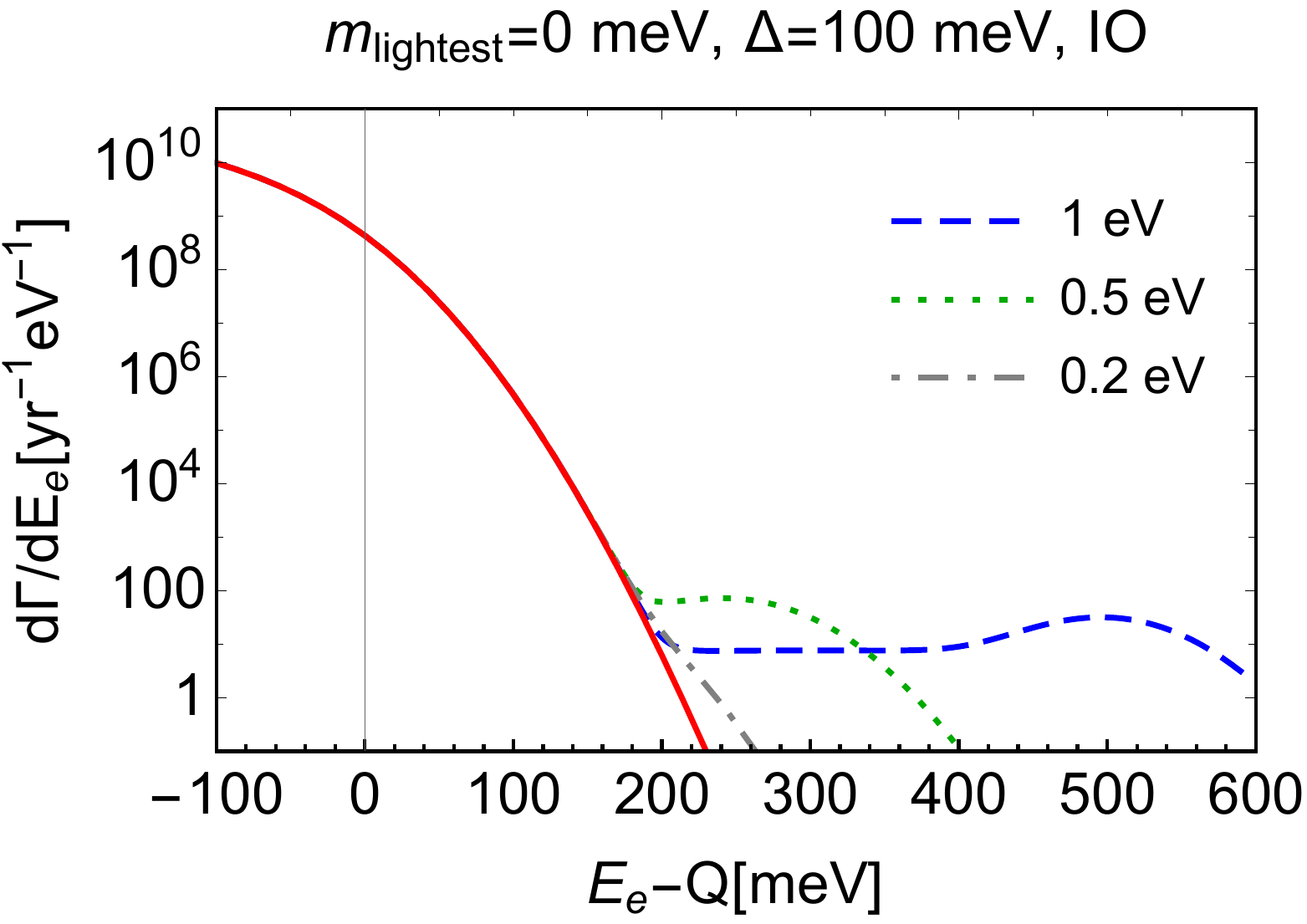}~\includegraphics[width=0.48\textwidth]{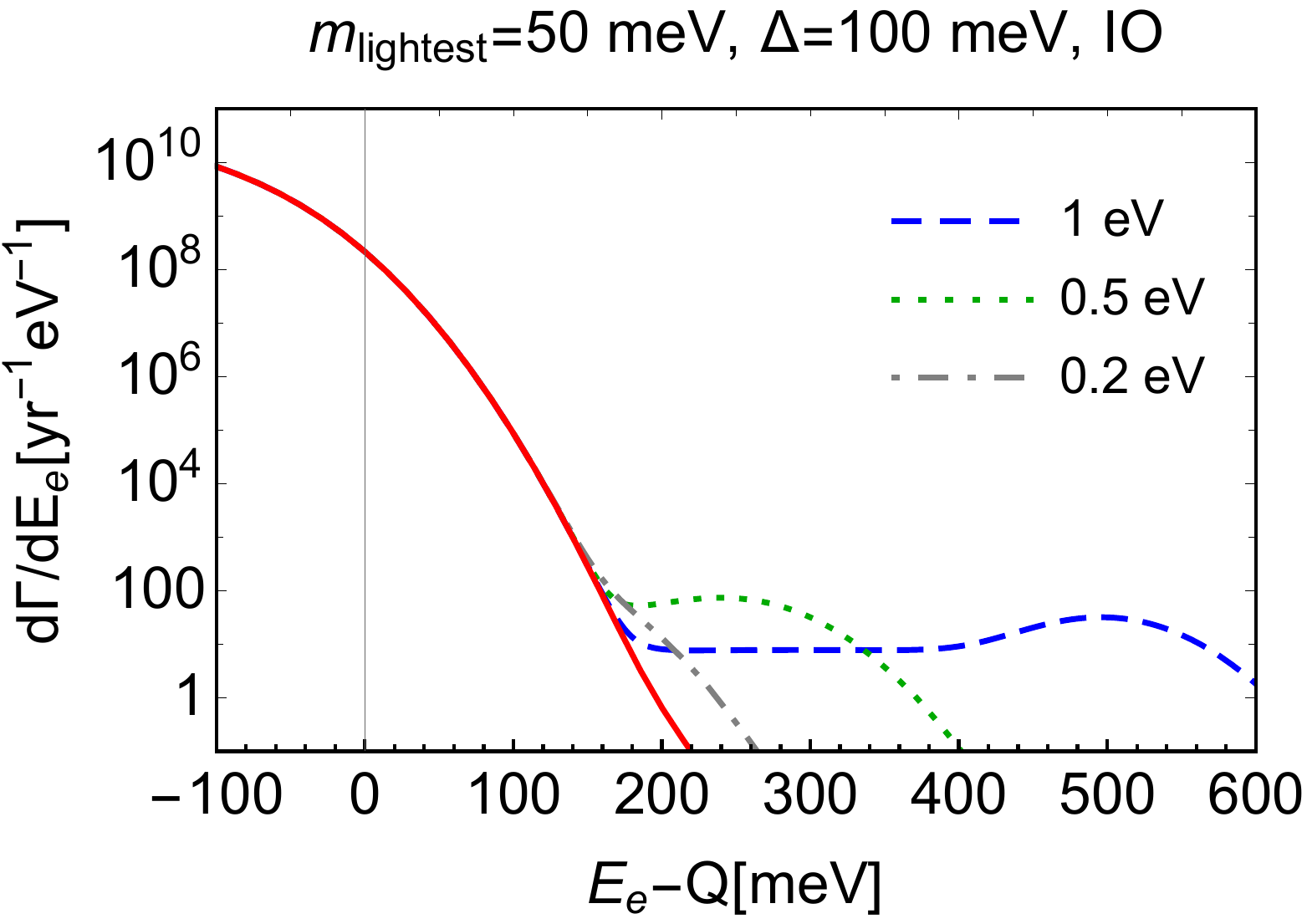}
    \caption{Examples of the DR neutrino signals from DM decay with $m_{\text{DM}}=1, 0.5$ and $0.2$~eV and inverse neutrino mass ordering (IO) with $m_{\nu_1}= 0$ (left panel) and $50$~meV (right panel). The DM lifetime is  taken as $\tau_{\text{DM}} = 10 t_{0}$ and the detector energy resolution is assumed to be $\Delta = 10$~meV, $50$~meV, and 100~meV from top to bottom.}
    \label{fig:electron-spectrum-IO}
\end{figure}

In this appendix Fig.~\ref{fig:electron-spectrum-IO} presents the pendant to Fig.~\ref{fig:electron-spectrum-NO} for an inverted neutrino mass ordering with two heavier states split by the smaller solar mass difference.

\section{Discovery potential for \boldmath$m_{\nu_1}=50~\mathrm{meV}$ ($m_{\nu_3}=50~\mathrm{meV}$)}
\begin{figure}[h]
    \centering
\includegraphics[width=0.5\textwidth]{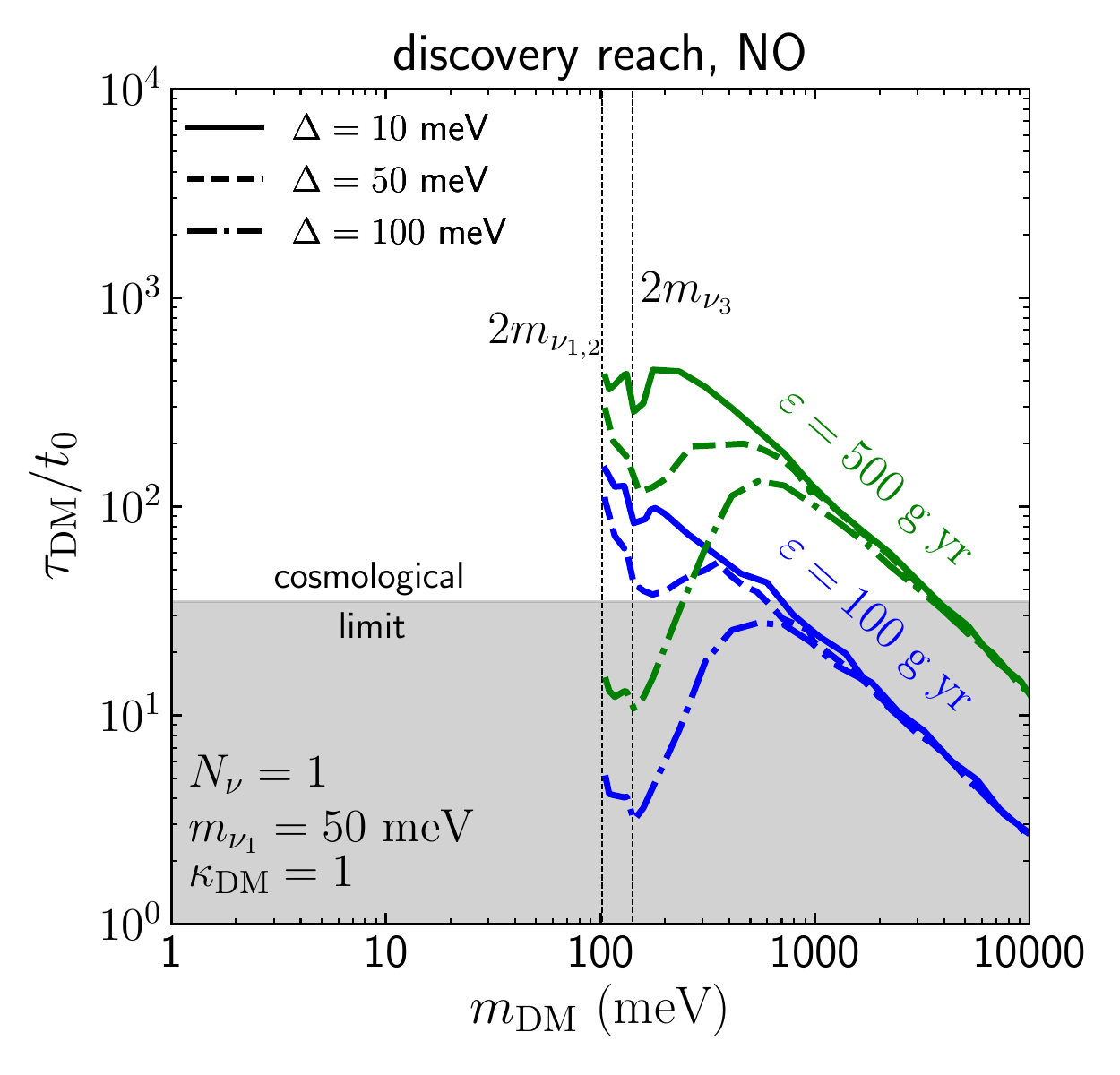}%
\includegraphics[width=0.5\textwidth]{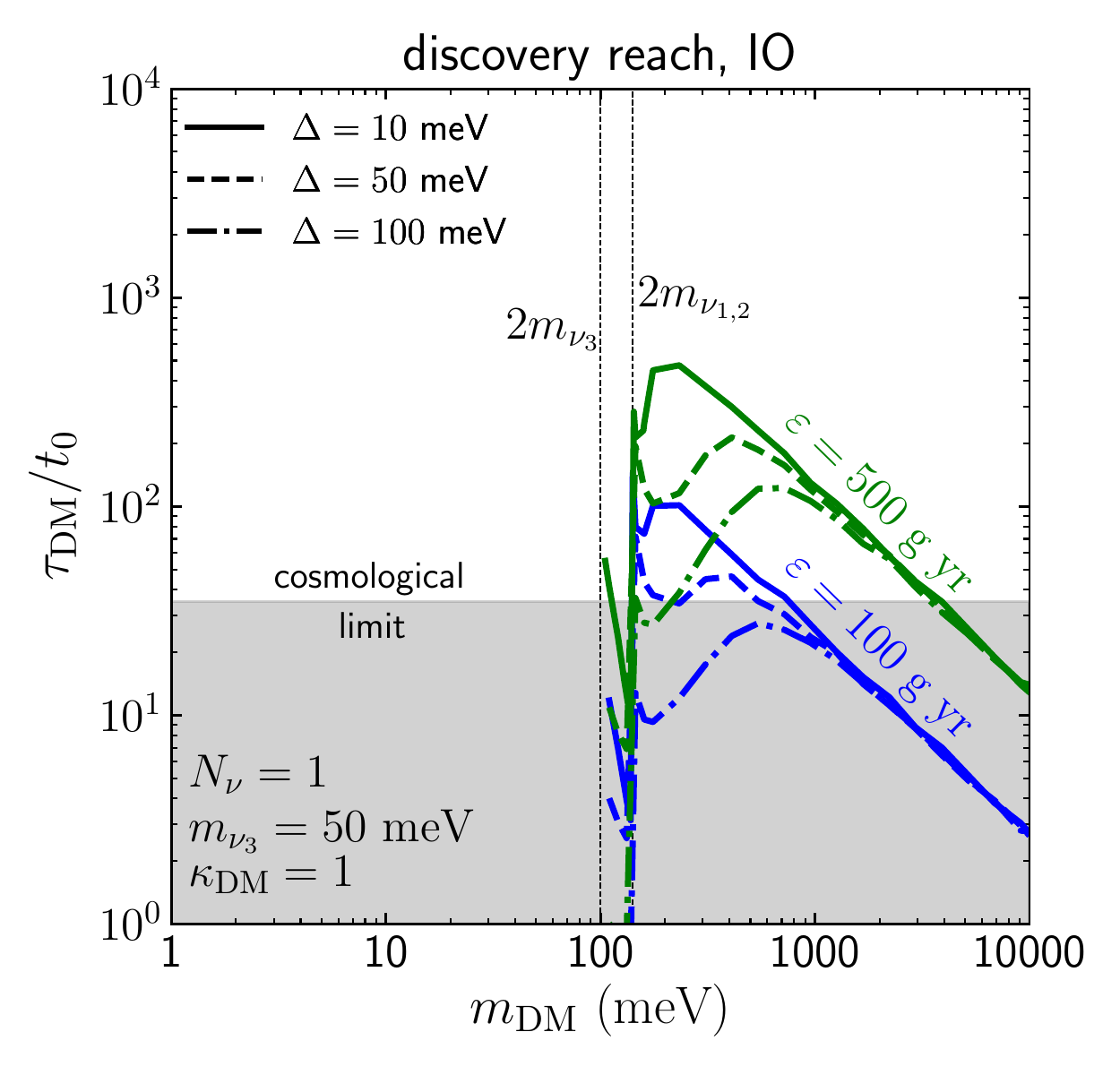}
    \caption{\small  Reach of PTOLEMY at $3\sigma$ significance as a function of progenitor mass $m_{\rm DM}$ and lifetime in units of the age of the Universe, $\tau_{\rm DM}/t_0$. An exposure of 100~g~yr (blue lines) and 500~g~yr (green lines) has been assumed for various projected performances on the electron energy resolution $\Delta$ as labeled, with  10~eV (100~eV) being the optimal (most conservative) case. All of DM is assumed to be decaying $\kappa_{\rm DM} = 1$. In the left (right) panel the mass of the lightest neutrino is $m_{\nu_1}=50~\mathrm{meV}$ ($m_{\nu_3}=50~\mathrm{meV}$) and a normal (inverted) hierarchy is assumed.} 
    \label{fig:limits_mnu50}
\end{figure}

\begin{figure}[h]
    \centering
\includegraphics[width=0.5\textwidth]{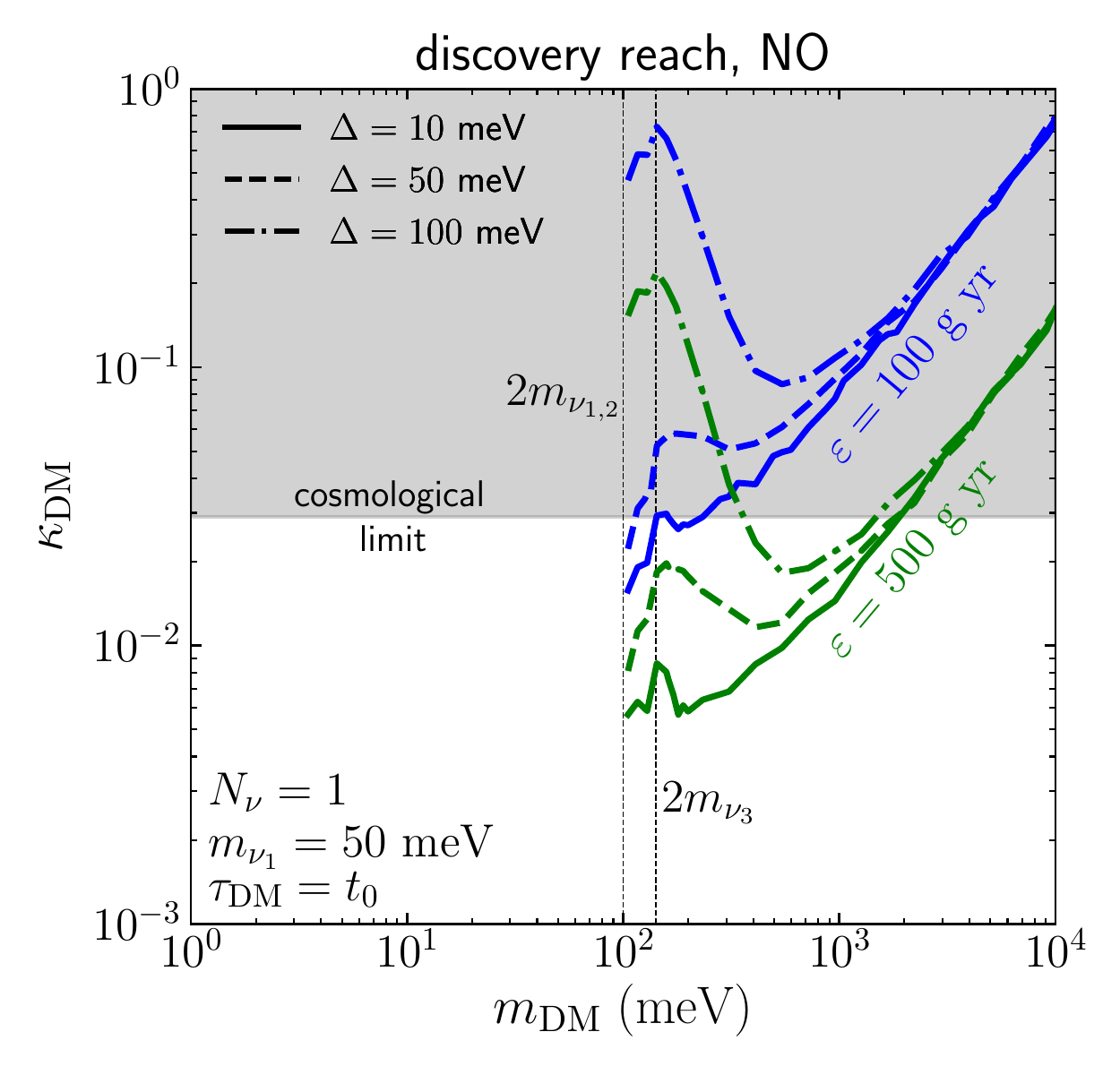}%
\includegraphics[width=0.5\textwidth]{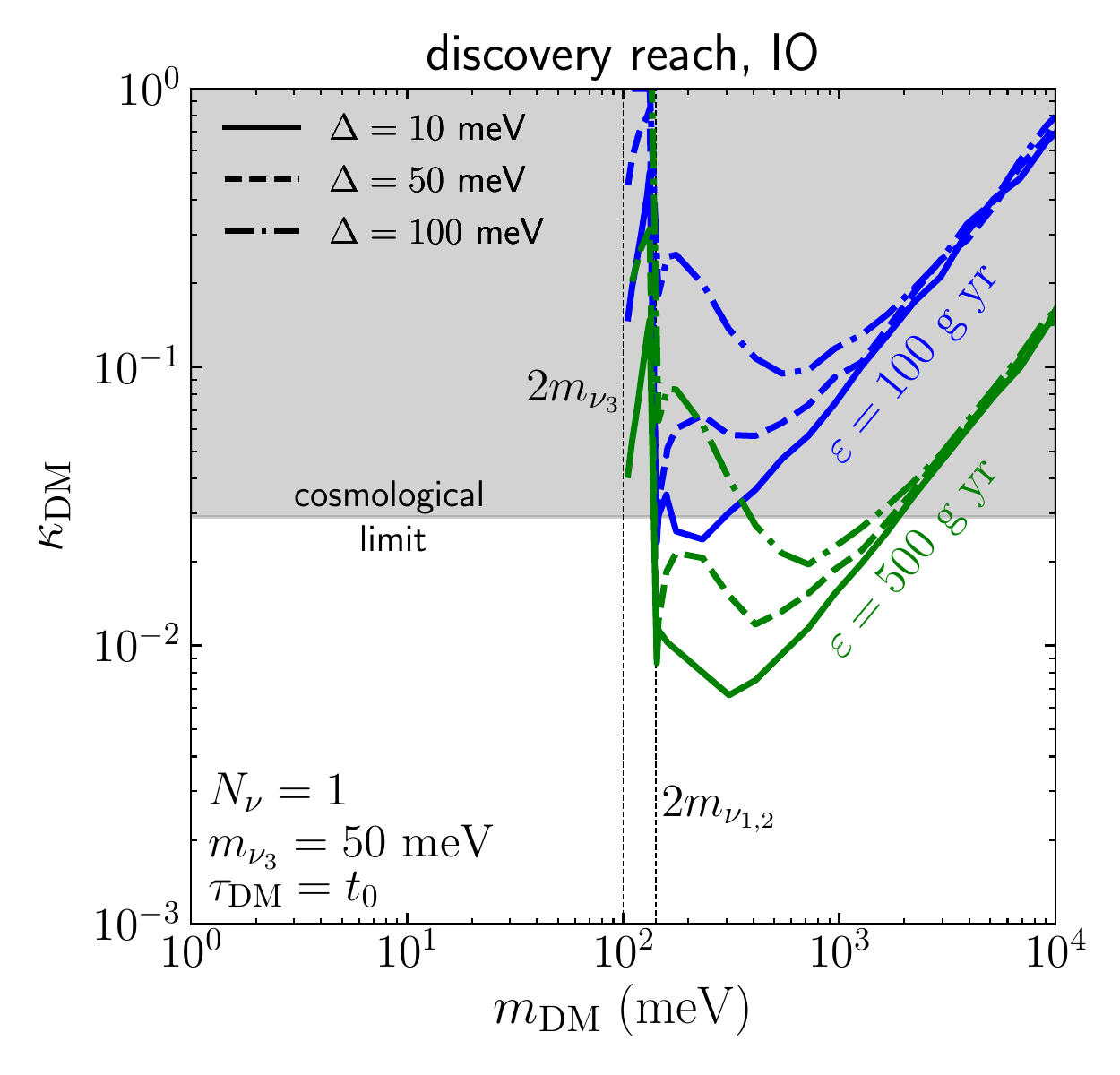}%
    \caption{\small Reach of PTOLEMY at $3\sigma$ significance to the decaying fraction of DM, $\kappa_{\rm DM}$, as a function of DM mass. The mass of the lightest neutrino is $m_{\nu_1}=50~\mathrm{meV}$ ($m_{\nu_3}=50~\mathrm{meV}$); a normal (inverted) hierarchy is assumed.}
    \label{fig:discFrac_mnu50}
\end{figure}

In Figs.~\ref{fig:limits_mnu50},~\ref{fig:discFrac_mnu50} we present the discovery potentials for $m_{\nu_1}=50~\mathrm{meV}$ ($m_{\nu_3}=50~\mathrm{meV}$) for normal (inverted) mass hierarchy. In this case, the minimum allowed DM mass is $100~\mathrm{meV}$ and the tritium beta background does not play a role. Hence, the $\CNB$ is the only background that enters in the analysis and only alters the limits around $m_\text{DM}=100~\mathrm{meV}$ compared to the figures~~\ref{fig:limits},~\ref{fig:discFrac}.

\bibliographystyle{JHEP}
\bibliography{ship.bib}

\end{document}